\documentclass[10pt]{article}
\usepackage{amsfonts,amssymb}
\usepackage{graphics,psboxit,amsmath} 
\usepackage{subfigure}
\usepackage{graphicx}
\usepackage{verbatim}
\usepackage{young}
\usepackage{array,multirow,hhline}
\usepackage{rotating}
\usepackage{tocbibind}

\def\hybrid{\topmargin -30pt    \oddsidemargin 0pt 
        \headheight 0pt \headsep 0pt
        \textwidth 6.25in       
        \textheight 9.5in       
        \marginparwidth .875in
        \parskip 5pt plus 1pt   \jot = 1.5ex}

\hybrid

\def\baselinestretch{1.2}

\catcode`\@=11

\def\marginnote#1{}
\newcount\hour
\newcount\minute
\newtoks\amorpm
\hour=\time\divide\hour by60
\minute=\time{\multiply\hour by60 \global\advance\minute by-\hour}
\edef\standardtime{{\ifnum\hour<12 \global\amorpm={am}%
        \else\global\amorpm={pm}\advance\hour by-12 \fi
        \ifnum\hour=0 \hour=12 \fi
        \number\hour:\ifnum\minute<10 0\fi\number\minute\the\amorpm}}
\edef\militarytime{\number\hour:\ifnum\minute<10 0\fi\number\minute}

\def\draftlabel#1{{\@bsphack\if@filesw {\let\thepage\relax
   \xdef\@gtempa{\write\@auxout{\string
      \newlabel{#1}{{\@currentlabel}{\thepage}}}}}\@gtempa
   \if@nobreak \ifvmode\nobreak\fi\fi\fi\@esphack}
        \gdef\@eqnlabel{#1}}
\def\@eqnlabel{}
\def\@vacuum{}
\def\draftmarginnote#1{\marginpar{\raggedright\scriptsize\tt#1}}

\def\draft{\oddsidemargin -.5truein
        \def\@oddfoot{\sl preliminary draft \hfil
        \rm\thepage\hfil\sl\today\quad\militarytime}
        \let\@evenfoot\@oddfoot \overfullrule 3pt
        \let\label=\draftlabel
        \let\marginnote=\draftmarginnote
   \def\@eqnnum{(\theequation)\rlap{\kern\marginparsep\tt\@eqnlabel}%
\global\let\@eqnlabel\@vacuum}  }

\def\draft2{
        \def\@oddfoot{\sl preliminary draft \hfil
        \rm\thepage\hfil\sl\today\quad\militarytime}
        \let\@evenfoot\@oddfoot \overfullrule 3pt
        \let\label=\draftlabel
        \let\marginnote=\draftmarginnote
   \def\@eqnnum{(\theequation)\rlap{\kern\marginparsep\tt\@eqnlabel}%
\global\let\@eqnlabel\@vacuum}  }

\def\preprint{\twocolumn\sloppy\flushbottom\parindent 2em
        \leftmargini 2em\leftmarginv .5em\leftmarginvi .5em
        \oddsidemargin -.5in    \evensidemargin -.5in
        \columnsep .4in \footheight 0pt
        \textwidth 10.in        \topmargin  -.4in
        \headheight 12pt \topskip .4in
        \textheight 6.9in \footskip 0pt
        \def\@oddhead{\thepage\hfil\addtocounter{page}{1}\thepage}
        \let\@evenhead\@oddhead \def\@oddfoot{} \def\@evenfoot{} }

\def\numberbysection{\@addtoreset{equation}{section}
        \def\theequation{\thesection.\arabic{equation}}}

\def\underline#1{\relax\ifmmode\@@underline#1\else
        $\@@underline{\hbox{#1}}$\relax\fi}

\def\titlepage{\@restonecolfalse\if@twocolumn\@restonecoltrue\onecolumn
     \else \newpage \fi \thispagestyle{empty}\c@page\z@
        \def\thefootnote{\fnsymbol{footnote}} }

\def\endtitlepage{\if@restonecol\twocolumn \else \newpage \fi
        \def\thefootnote{\arabic{footnote}}
        \setcounter{footnote}{0}}  

\catcode`@=12
\relax

\def\figcap{\section*{Figure Captions\markboth
        {FIGURECAPTIONS}{FIGURECAPTIONS}}\list
        {Figure \arabic{enumi}:\hfill}{\settowidth\labelwidth{Figure
999:}
        \leftmargin\labelwidth
        \advance\leftmargin\labelsep\usecounter{enumi}}}
 \relax
\def\tablecap{\section*{Table Captions\markboth
        {TABLECAPTIONS}{TABLECAPTIONS}}\list
        {Table \arabic{enumi}:\hfill}{\settowidth\labelwidth{Table
999:}
        \leftmargin\labelwidth
        \advance\leftmargin\labelsep\usecounter{enumi}}}
 \relax
\def\reflist{\section*{References\markboth
        {REFLIST}{REFLIST}}\list
        {[\arabic{enumi}]\hfill}{\settowidth\labelwidth{[999]}
        \leftmargin\labelwidth
        \advance\leftmargin\labelsep\usecounter{enumi}}}
 \relax
\makeatletter
\newcounter{pubctr}
\def\publist{\@ifnextchar[{\@publist}{\@@publist}}
\def\@publist[#1]{\list
        {[\arabic{pubctr}]\hfill}{\settowidth\labelwidth{[999]}
        \leftmargin\labelwidth
        \advance\leftmargin\labelsep
        \@nmbrlisttrue\def\@listctr{pubctr}
        \setcounter{pubctr}{#1}\addtocounter{pubctr}{-1}}}
\def\@@publist{\list
        {[\arabic{pubctr}]\hfill}{\settowidth\labelwidth{[999]}
        \leftmargin\labelwidth
        \advance\leftmargin\labelsep
        \@nmbrlisttrue\def\@listctr{pubctr}}}
 \relax
\makeatother

\def\be{\begin{equation}}
\def\ee{\end{equation}}
\def\ba{\begin{eqnarray}}
\def\ea{\end{eqnarray}}

\def\a{\alpha}

\def\no{\noindent}

\def\qq{\qquad}

\def\IR{\relax{\rm I\kern-.18em R}}

\def\inv{^{\raise.0ex\hbox{${\scriptscriptstyle -}$}\kern-.05em 1}}

\def\const{{\rm const.}}

\begin{document}


\renewcommand{\theequation}{\thesection.\arabic{equation}}
\csname @addtoreset\endcsname{equation}{section}

\begin{titlepage}
\begin{center}

\begin{flushright}
 \hfill AGIF-MRGC 2/2013
\end{flushright}

\phantom{xx}
\vskip 0.5in

{\large \bf Solutions of massive gravity theories in constant scalar invariant geometries }

\vskip 0.5in

{\bf K. Siampos}${}^{a}$,\phantom{x} {\bf Ph. Spindel}${}^{b}$\phantom{x}
\vskip 0.1in

M\'ecanique et Gravitation, Universit\'e de Mons,
7000 Mons, Belgique.

\vskip .2in

\end{center}

\vskip .4in

\centerline{{\bf Synopsis}}

We solve massive gravity field equations in the framework of locally homogenous and vanishing scalar invariant (VSI) Lorentzian  spacetimes,
which in three dimensions are the building blocks of constant scalar invariant (CSI)  spacetimes.
At first, we provide an exhaustive list of all Lorentzian three-dimensional homogeneous spaces and then we determine the Petrov type of the relevant curvature tensors. 
Among these geometries we determine for which values of their structure constants they are solutions of the field equations of massive gravity theories 
with cosmological constant. The homogeneous solutions founded are of all various Petrov types : 
$I_{\mathbb C}$, $I_{\mathbb R}$, $II$, $III$, $D_t$, $D_s$, $N$, $O$; the VSI geometries which we found are of Petrov type $III$.  
The Petrov types $II$ and $III$ are new explicit CSI  spacetimes solutions of these types.
We also examine the conditions under which the obtained anti-de Sitter solutions   are free of tachyonic massive graviton modes.

\vfill
\no
 {
 $^a$konstantinos.siampos@umons.ac.be,\phantom{x}
 $^b$philippe.spindel@umons.ac.be}

\end{titlepage}
\vfill
\eject


\tableofcontents

\def\baselinestretch{1.2}
\baselineskip 20 pt
\no

\newcommand{\eqn}[1]{(\ref{#1})}

\section{Introduction}

It is known that, in three dimensions, Einstein gravity theory does not possess any local physical degrees of freedom. 
However Deser, Jackiw and Templeton 
proposed a modification of the theory \cite{DJH}, consisting of a Chern--Simons term \cite{DJT1, DJT2} added to the usual Einstein--Hilbert Lagrangian. The latter is known 
as topological massive gravity (TMG). This is a chiral three-dimensional gravity theory which contains massive spin-2 excitations that mediate finite-range interactions. Moreover, this modification
provides an interesting framework for a unitary quantum theory of gravity.\footnote{Its status concerning renormalisability seems not yet definitely established \cite{Deser:1990bj, Nakasone:2009bn,Deser:2009hb,Oda:2009ys}. We thank S. Deser for a comment on this point.}

On the other hand Bergshoeff, Hohm and Townsend \cite{BHT1,BHT2} have recently obtained another type of massive gravity theory (NMG): a 
parity preserving theory that describes (on a Minkowski background) the propagation of a massive positive energy spin-2 field, but now of 
both helicities : $\pm2$. This theory possesses all the 
virtues of the TMG, so it may be considered as another consistent candidate of a theory of 3D quantum gravity. 
These authors have also considered the ``merging" of both theories, producing a general massive gravity theory  (GMG), which involves two spin-2 helicity states with different masses (parity-violating) and, as the previous ones, can also be  extended by adding a cosmological constant. 
{\par\noindent Let us mention that solutions of the NMG theory, including $AdS_3$ black holes, were constructed in \cite{NMGSol}. In addition, aspects of the gauge/gravity duality in this theory were found to be in agreement with the holographic studies of TMG theory \cite{NMG.AdS, Anninos:2008fx}.}

This work is motivated by the one of Chow et al. \cite{CPS1}, which provided a review of a large set of solutions of topological 
massive gravity (with cosmological constant). 
In their paper these authors described a three-dimensional variant of Petrov classification and showed that all the solutions that were founded in the literature at this 
time were of Petrov types $D$ or $N$, corresponding to locally squashed $AdS_3$ or $AdS$ pp-wave metrics. Moreover they proved that all 
Petrov type $D$ solutions of TMG actually are biaxially squashed $AdS_3$ metrics.
In a companion paper \cite{CPS2} these authors also obtained new solutions of the topologically massive gravity equations by considering 
Kundt metrics \cite{Ku,KT} 
(see also Chapters 28 and  31 of \cite{SKMHH}). The TMG solutions belonging to this class of metrics are generically of Petrov type $II$, but there are some special 
cases of Petrov types $D$, $III$, $N$ and $O$ as well. Let us notice that a classification of the homogeneous solutions  of TMG equations (without cosmological constant) was 
given by Ortiz  \cite{Ortiz1} and  was recently generalised for non-vanishing cosmological constant by Moutsopoulos  \cite{Moutsopoulos:2012bi}.

On the other hand, Coley et al. have proved  \cite{Coley:2007ib}    that a Lorentzian three-dimensional spacetime  on which all scalars built out of the curvature tensor are constant ($CSI$ spaces), 
can be constructed by means of fibering and warping, from locally homogeneous spaces and a subclass of CSI spaces, namely the vanishing scalar 
invariant spaces ($VSI$). 

{The purpose of the present work is to obtain all scalar invariant geometries (i.e. locally homogenous and VSI Lorentzian  spacetimes) that solve the massive gravity (MG) theory equations with cosmological constant and to classify them according to their Petrov types.}

The paper is organised as follows: In section \ref{MGeqs}  we summarise the various field equations of the massive gravity theories. 
We start section \ref{Hom.geom} by revisiting all the three-dimensional homogeneous spaces with Lorentzian signature and determine
the Petrov types of the relevant curvature tensors (listed in the appendix), which will prove to be useful for
solving the equations and identifying the solutions.
We then proceed in section \ref{Hom.Sol} to solve the equations of the massive gravity theories  on homogeneous spaces. Finally, in section \ref{VSI} we study massive gravity equations
on VSI spaces.

\section{Massive Gravity Theories}
\label{MGeqs}

{This section is devoted to a short review of the various massive gravity theories in three dimensions.}

They all consist by adding extra pieces to the usual Einstein--Hilbert action :
\begin{eqnarray}
\label{EH action}
S_{\rm EH}=\frac{1}{\kappa^2}\int \sqrt{\vert g\vert}(R-2\Lambda)\,\mathrm{d}^3x\qquad,\qquad
g:=\det(g_{\mu\,\nu})
\end{eqnarray}
where $\Lambda$ is the cosmological constant and $\kappa$ is the gravitational coupling with
mass dimension $[\kappa]=-\frac{1}{2}$\,. In what follows we adopt the mostly plus expression of the metric and
define the curvature tensors so that the curvature of the Euclidean round sphere, equipped with its positive definite 
metric, has positive curvature\footnote{In others words according to the conventions : 
$R^\mu_{\ \nu\,\rho\,\sigma}=\partial_\rho\,\Gamma^{\mu}_{\nu\,\sigma}+\dots $; 
$R_{\nu\,\sigma}=R^\rho_{\ \nu\,\rho\,\sigma}~. $} and fix the spacetime orientation by adopting for the tensor density $\varepsilon_{1\,2\,3}=+1$ (so $\varepsilon^{1\,2\,3}=g^{-1}$) .
\subsection{ Topologically massive gravity}
 Topological massive gravity is obtained by adding a gravitational Chern--Simons
term to the Einstein--Hilbert action
\begin{eqnarray}
\label{TMG} 
&&S_{\rm TMG }: =S_{\rm EH}+\frac{1}{\mu\,\kappa^2}\,S_{\rm CS}\qquad,\\
&&S_{\rm CS}=\frac{1}{2}\int g\,\varepsilon^{\lambda\mu\nu}\,\Gamma_{\lambda\sigma}{}^\rho
\left(\partial_\mu\Gamma_{\rho\nu}{}^\sigma+\frac{2}{3}\Gamma_{\mu\tau}{}^\sigma\Gamma_{\nu\rho}{}^\tau\right)\,\mathrm{d}^3x\qquad, \nonumber
\end{eqnarray}
which is expressed through Christoffel symbols of the spacetime metric $g_{\mu\nu}$, while $\mu$ is a new coupling constant with mass dimension
one. \\
The classical equations of motion read as
\begin{eqnarray}
\label{eomTMG} 
&&R_{\mu\nu}-\frac{1}{2}g_{\mu\nu}R+\Lambda g_{\mu\nu}+\frac{1}{\mu}\,C_{\mu\nu}=0\qquad,\\
&&C_{\mu\nu}:=\sqrt{|g|}\,\varepsilon_{\mu\rho\sigma}\nabla^\rho\left(R_\nu{}^\sigma-
\frac{1}{4}\delta_\nu{}^\sigma\,R\right)\qquad,\nonumber
\end{eqnarray}
where $C_{\mu\nu}$ is the Cotton--York tensor :  a symmetric, traceless and divergenceless tensor.
An immediate consequence of the equations of motion is that the traceless part of the Ricci tensor and the Cotton--York are proportional
\begin{equation}
\label{eomTMGtr} 
S_{\mu\nu}+\frac{1}{\mu}\,C_{\mu\nu}=0\qquad,\qquad
S_{\mu\nu}:=R_{\mu\nu}-\frac{1}{3}g_{\mu\nu}R\qquad,
\end{equation}
 and accordingly of the same Petrov type.
\subsection{New massive gravity}
This theory is defined by adding a quadratic curvature term \cite{BHT1} to the Einstein--Hilbert action
\begin{eqnarray}
\label{NMG} 
&&S_{\rm NMG}: =S_{\rm EH}-\frac{1}{\xi\,\kappa^2}\,S_{\rm QC}\qquad,\\
\label{QC}
&&S_{\rm QC}=\int \sqrt{\vert g\vert}\left(R_{\mu\nu}R^{\mu\nu}-\frac{3}{8}R^2\right)\,\mathrm{d}^3x\qquad,\nonumber
\end{eqnarray}
where $\xi$ is a coupling constant with mass dimension two. The corresponding equations of motion read
\begin{eqnarray}
\label{eomNMG} 
&&R_{\mu\nu}-\frac{1}{2}g_{\mu\nu}R+\Lambda g_{\mu\nu}-\frac{1}{2\xi\,}\,K_{\mu\nu}=0\qquad,\\
&&K_{\mu\nu}=2\nabla^2R_{\mu\nu}-\frac{1}{2}\nabla_\mu\nabla_\nu R+\frac{9}{2}R\,R_{\mu\nu}-8R_\mu{}^\kappa R_{\nu\kappa}
+g_{\mu\nu}\left(3R_{\kappa\lambda}R^{\kappa\lambda}-\frac{1}{2}\nabla^2R-\frac{13}{8}R^2\right)\qquad, \nonumber
\end{eqnarray}
where $K_{\mu\nu}$ is a symmetric and divergenceless tensor.  
 Similarly to Eq.\eqn{eomTMGtr} we see that the traceless parts of the Ricci tensor and $K_{\mu\nu}$ tensor are proportional
\begin{equation}
\label{eomNMGtr} 
S_{\mu\nu}=\frac{1}{2\xi\,}\widehat K_{\mu\nu}\qquad,\qquad \widehat K_{\mu\nu}:=K_{\mu\nu}-\frac{1}{3}g_{\mu\nu}K\qquad,
\end{equation}
and thus of the same Petrov type.
 
\subsection{General Massive Gravity}
It is defined by a combination of the topological and the new massive gravity theories
\begin{eqnarray}
\label{GMG}
S_{\rm GMG}:=S_{\rm EH}+\frac{1}{\mu\,\kappa^2}~S_{\rm CS}-\frac{1}{\xi\,\kappa^2}~S_{\rm QC}\qquad,
\end{eqnarray}
and the corresponding field equations read
\begin{eqnarray}
R_{\mu\nu}-\frac{1}{2}g_{\mu\nu}R+\Lambda g_{\mu\nu}+\frac{1}{\mu}\,C_{\mu\nu}-\frac{1}{2\xi\,}\,K_{\mu\nu}=0\qquad.
\end{eqnarray}
To identify the field content of the latter equations, we have to linearise them around a fixed background.
This was performed around Minkowsky and anti-de Sitter ($AdS_3$) backgrounds in \cite{BHT1} and \cite{Liu:2009pha} respectively. 
In particular, it was shown in \cite{Liu:2009pha} that there are two massive spin-2 modes, 
which are stable if
\begin{equation}\label{stabcond}
\xi(\xi+4\mu^2)\geqslant2\Lambda\mu^2\qquad.
\end{equation}
Thus, for pure NMG, when $\mu$ goes to infinity,  an $AdS_3$ space is exempt of tachyonic graviton if $\xi\geqslant\frac{\Lambda}{2}$. 
In what follows we shall not restrict the sign of the coupling constants, but unless for exact $AdS_3$ geometries, just provide some plots of their signs according to the parameters appearing 
in the expressions of the metrics we obtain. Indeed, for an asymptotically anti-de Sitter space ($\Lambda<0$), condition \eqn{stabcond} implies that  for $\xi>0$ the graviton has 
no tachyonic massive mode.
\section{Homogeneous geometries}
\label{Hom.geom}

{In this section we review the various expressions of the structure constants of the isometry groups characterising  the three dimensional locally  homogenous space times which are going to be used in section \ref{Sol.trans} for solving the  MG field equations.}

The strategy we adopt to obtain all homogeneous space metrics, that are solutions of the massive gravity field equations, was
the one advocated a long time ago by Ozsv\'ath \cite{Oz} (see also \cite{Ortiz1,Moutsopoulos:2012bi}). Let us briefly remind its principle. We consider metrics on homogeneous 
three-dimensional spaces invariant under the action of a locally simply transitive isometry group. The action  being simply transitive implies that 
locally such spaces can be identified with the groups acting on them. Suppose that we choose the left action of 
$G$ on itself. The vectors tangent to the orbits of the one-parameter subgroups of $G$ constitute right invariant vector fields obeying the relations  
\begin{equation}\label{Cartan}
[\xi_\alpha,\xi_\beta]=-C^\gamma_{\alpha\,\beta}\,\xi_\gamma\qquad,
\end{equation}
where $C^\gamma_{\alpha\,\beta}$ are the structure constants of the Lie algebra of $G$. The group being simply transitive also 
implies that at each point these vectors $\xi_\alpha$ constitute a local frame. Their dual 1-form $\theta^\alpha$ (such that $\theta^\alpha(\xi_\beta))=\delta^\alpha_\beta$)
define the right invariant coframes. Metric tensors whose components with respect to these coframes are constants, also are right invariant and the generators $\tilde\xi_\alpha$ associated to the right action of $G$ are their Killing vector fields.
In the framework of three-dimensional groups, it is well known how to implement the action of $GL(3,\mathbb{R})$ and writing  $C^\gamma_{\alpha\,\beta}$
in a canonical form according to the  Bianchi classification \cite{SKMHH}. However, solving the gravitational field equations in a Lorentzian invariant theory
appears to be much easier by setting the metric in a canonical form :
\begin{equation}
(\eta_{\alpha\,\beta})=diag. (-1,1,1)\,,\label{canmet}
\end{equation}
but starting from arbitrary structure constants of the Lie algebra  and fixing them by $Iso(1,2)$ transformations.  In this framework, all geometrical quantities expressed in the invariant coframe are 
algebraic functions of the structure constants. For instance, defining:  
$
C_{\alpha\,\beta\,\gamma}:= C_{\alpha\,\beta}^\delta\,\eta_{\delta\,\gamma}\  ,
$
we obtain the connection coefficients, the Ricci tensor and its covariant derivative~: 
\begin{equation}
\omega_{\alpha\,\beta\,\gamma}= \frac 12(C_{\beta\,\gamma\,\alpha}-C_{\alpha\,\gamma\,\beta}-C_{\alpha\,\beta\,\gamma})\ ,
R_{\alpha\,\beta}=\omega^\gamma_{\phantom{\mu\,}\delta\,\gamma}\,\omega^\delta_{\phantom{\nu\,}\alpha\,\beta}-
\omega^\gamma_{\phantom{\mu\,}\alpha\,\delta}\,\omega^{\delta}_{\phantom{\nu\,}\beta\,\gamma}\  ,
\nabla_\gamma\,R_{\alpha\,\beta}=-R_{\delta\,\beta}\,\omega^{\delta}_{\phantom{\mu\,}\alpha\,\gamma}-
R_{\alpha\,\delta}\,\omega^{\delta}_{\phantom{\mu\,}\beta\,\gamma}\,.
\end{equation}
Accordingly all the field equations, we shall encounter in section \ref{Hom.Sol}, become algebraic equations.
But first we shall review the classification of the structure constants, then put them into the field equations and solve the resulting algebraic equations.

\subsection{Bianchi classification revisited}
As it was mentioned above, the Lie algebras of the (right) invariant fields ( given by Eq.\eqn{Cartan},
whose structure constants satisfy the Jacobi identity 
$C^\delta_{\kappa\,\alpha}\,C^\kappa_{\beta\,\gamma}\,\varepsilon^{\alpha\,\beta\,\gamma}=0$) allocate into two classes. 
The unimodular Lie algebras, such that $C^\kappa_{\alpha\kappa\,}=0$ and the non-unimodular ones such that
$\frac 12 C^\kappa_{\alpha\kappa\,}=k_\alpha\neq 0$. 
For the three-dimensional real algebras (first classified by Bianchi), their structure constants can be parametrised in 
terms of the vector components $k_a$ and a symmetric tensor density\footnote{In what follows we shall call them respectively 
{\it structure vector} and {\it structure tensor density}.} 
$n^{\alpha\,\beta} $ (see \cite{SKMHH} and refs therein),
\begin{equation}\label{Cabc}
C^\alpha_{\beta\,\gamma}=\varepsilon_{\beta\,\gamma\,\zeta}n^{\zeta\,\alpha}+k_\beta\,\delta^\alpha_\gamma-k_\gamma\,\delta^\alpha_\beta\qquad ,
\end{equation}
while the Jacobi identity reduces to the condition $k_\alpha\,n^{\alpha\,\beta}=0$.
Thus, the $Iso(1,2)$ classification of the structure constants reduces 
to the classification of symmetric tensor densities  that annihilate  a vector, a problem that was solved in \cite{Pleb}  (see also refs \cite{Hiromoto:1978qb, HwEl}).\\
Chow et al. \cite{CPS1} suggested to classify solutions of TMG according to the  Segre classification of the traceless Ricci 
tensor $S_{\alpha\,\beta}$, which, in their framework, is equivalent to the Petrov classification of the Cotton--York tensor but not 
necessarily in the case of NMG. Hereafter, we shall present the various canonical forms of the Lie algebras obtained using $Iso(1,2)$ transformations. We also determine
 the Segre--Petrov types of the traceless Ricci, Cotton--York and $\widehat K_{\alpha\,\beta}$ tensors obtained from their invariant coframes and the canonical form (\ref{canmet}) of the metric . On a practical level, to obtain this classification it is not necessary
to know explicitly the eigenvalues of the tensor, when they all are different. So, we have just to compute the discriminant of its 
characteristic equation. If it is  positive, then two eigenvalues are complex conjugate and one is real : it corresponds to the case 
$I_{\mathbb C}$ (see ref.\cite{CPS1} for the notations). If it is negative the three eigenvalues are real and distinct, corresponding 
to the case $I_{\mathbb{R}}$.  It is only when it is zero, in which case we know that at least one eigenvalue is double, that further 
analysis is required to determine the Jordan form of the tensor. In other words, it is the degeneracy of the eigenvalues that greatly facilitates the analysis. 
In this case, when the characteristic polynomial reduces to its cubic term $P(\lambda)=\lambda^3$, the tensor is of Petrov type 
$O$, $N$, or $III$; otherwise if the three roots are equal, but non vanishing, 
the tensor is of Petrov type $D_s$, $D_t$ or $II$.

\subsection{Unimodular Lie algebras}

Four types of normal forms are possible~:
\begin{itemize}
\item{Type $I$}  
\begin{equation}\label{nI}
(n_I^{\alpha\,\beta})= \left( \begin{array}{ccc}
    a & 0 & 0 \\ 
   0 & b & c \\ 
    0 & c & b \\ 
  \end{array}\right)\qquad ,
\end{equation}
this form is equivalent to $diag.\, (a,b+c,b-c)$, but the latter appears to be more easy to handle for solving the field equations. 
By changing the orientation (in which case $\mu$ is changing sign), we may always assume that if $a$ is non zero it is positive, otherwise 
that $b>0$ if $b$ is non zero, etc.\\ 
Obviously this structure tensor density may correspond to any of the unimodular Bianchi types ($I,\,II,\, VI_0,\,VII_0,\, VIII,\,IX $) 
of real Lie algebras (see for instance Ortiz \cite{Ortiz1}).\\
The traceless part of the Ricci tensor is :
\begin{equation}\label{SnI}
\left( S_{\alpha\,\beta}\right)=\left(
\begin{array}{ccc}
 \frac{2}{3} \left(a^2+a\,b-2 c^2\right) & 0 & 0 \\
 0 & \frac{1}{3} \left(a^2+a\,b-2 c^2\right) & -(a+2 b) c \\
 0 & -(a+2 b) c & \frac{1}{3} \left(a^2+a\,b-2 c^2\right)
\end{array}
\right)\qquad .
\end{equation}
To determine its Petrov type, we have to obtain the Jordan form of the matrix of components $S^\alpha_\beta$. 
Fortunately, as already mentioned, we will not have to handle explicit solutions of the third degree characteristic polynomial 
\begin{equation}
P(\lambda):= \det [\lambda\, \delta^\alpha_\beta- S^\alpha_\beta]
\end{equation}
in the general case. The traceless condition implies that this cubic polynomial will always be of the form : $P(\lambda)=\lambda^3+p\,\lambda+q$. Its discriminant, which is defined as 
$\Delta:= \frac{q^2}4+\frac{p^3}{27}$, partially fixes the number and the nature of its different roots. If    $\Delta >0$ one root is real and the two others are complex conjugate; if $\Delta<0$ the three roots are real and distinct; if $\Delta=0$, at least two roots are equal. Accordingly, when $\Delta>0$ the traceless tensor will be of Petrov type $I_{\mathbb C}$ and when $\Delta<0$ of Petrov type $I_{\mathbb R}$. When $\Delta =0$, the tensor is of special Petrov type. It is of type $II$ or $D$ when $p\not =0$, and of type $III$, $N$ or $O$ when $P(\lambda)=\lambda^3$. Then its precise determination will need more investigation, but things are greatly facilitated because at least one of the eigenvalues of the tensor is degenerate. 
For example, in case of the   tensor    \eqn{SnI}, we obtain
\begin{equation}
\Delta_{\rm S}=-\frac1{27}\,(a+2\,b)^2\,c^2\,\left((a+b)^2-c^2\right)^2\,(a^2-4\,c^2)^2~,
\end{equation}
which generically is negative and thus define a tensor of type $I_{\mathbb R}$; exceptions occur when it vanishes.  
The results of this analysis, both for the 
$S_{\alpha\,\beta}$ with $C_{\alpha\,\beta}$ and
$\widehat K_{\alpha\,\beta}$
tensors, are summarised in tables 1-6.
Thus, solutions of TMG or NMG can be easily found by requiring matching of the Petrov classifications of $S_{\alpha\,\beta}$ with $C_{\alpha\,\beta}$ or 
$\widehat K_{\alpha\,\beta}$
respectively. However, this is not {\it  a priori} the case for GMG.

\item{Type $II$} 
\begin{equation}
\label{tensor.II}
\qquad(n_{II}^{\alpha\,\beta})= \left( \begin{array}{ccc}
    \nu+a & \nu & 0 \\ 
   \nu & \nu-a & 0 \\ 
    0 & 0 & b \\ 
  \end{array}\right)\quad , \quad \nu=\pm 1\qquad ,
\end{equation}
which does not admit timelike eigenvector but a double null vector.\\
Diagonalising $n_{II}$ with a $GL(3,\mathbb{R})$ transformation, we see that if $a=b=0$, it corresponds 
to a Lie algebras of Bianchi  type $II$, if $b=0$ but $a\neq 0$ or $a=0$ and $\nu\,b< 0$ to the ones of Bianchi type $VI_0$, 
if $a=0$ and $\nu\,b>0$ to Bianchi type $VII_0$, and otherwise to Bianchi type $VIII$.

\item{Type $III$}  
\begin{equation}\label{nIII}
(n_{III}^{\alpha\,\beta})= \left( \begin{array}{ccc}
    a & 1 & 0 \\ 
   1 & -a & 1  \\ 
    0 & 1 & -a  \\ 
  \end{array}\right)\qquad,
\end{equation}
which admits a triple null vector. \\
This structure constant density may only correspond to Lie algebras of Bianchi type $VI_0$ if $a=0$
or $VIII$ if  $a\neq0$.

\item{Type $IV$} 
\begin{equation}
(n_{IV}^{\alpha\,\beta})= \left( \begin{array}{ccc}
    0 & \nu & 0 \\ 
   \nu & a & 0  \\ 
    0 & 0 & b  \\ 
  \end{array}\right)\qquad,
\end{equation}
where $a^2<4\,\nu^2$ and that has only one simple spacelike eigenvector, but no timelike or null eigenvector.\\
Here again, only the structure constants of Lie algebras of Bianchi types $VI_0$ if $b=0$ or $VIII$ if $b\neq0$ are available.

\end{itemize}
\subsection{Non-unimodular Lie algebras}
In the case of non-unimodular Lie algebras, we have to consider three possibilities: the vector 
$k_\alpha$ is timelike, spacelike or null; four type of normal forms will occur. 
\begin{itemize}
\item {Timelike $k_\alpha$ : }We  choose the frame such that $k_\alpha=(k,0,0)$. The Jacobi identity implies that 
$n^{\alpha\,\beta}$ is a spacelike symmetric tensor that can be diagonalised by a rotation in the $[1,\,2]$ plane. 
But  again this normal form turns out not to be  the most suitable one for solving the field equations, and we prefer to use the following 
form
\begin{equation}
(n_T^{\alpha\,\beta})= \left( \begin{array}{ccc}
    0 & 0 & 0 \\ 
   0 & a & b \\ 
    0 & b & a \\ 
  \end{array}\right)\qquad \qquad \mbox{(Type $T$)}\qquad .
\end{equation}
Obviously, all non-unimodular Lie algebras may lead to this form of the structure tensor density. 
More precisely we have  Lie algebras of Bianchi type $V$ if $a=b=0$, of type $IV$ if $a=\pm b\neq 0$, and 
otherwise of types $III$, $VI_h$ for $\vert b\vert>\vert a\vert$ or $VII_h$ for $\vert a\vert>\vert b\vert$ with $h=k^2/(a^2-b^2)$.

\item {Spacelike $k_\alpha$ : } 
We  choose the frame such that $k_\alpha=(0,0,k)$. Then the structure tensor density may take three different canonical 
forms. If it  admits a  timelike  (and thus a spacelike) eigenvector 
\begin{equation}
(n_{S\,I}^{\alpha\,\beta})= \left( \begin{array}{ccc}
    a & 0 & 0 \\ 
   0 & b & 0 \\ 
    0 & 0 & 0 \\ 
  \end{array}\right)\qquad \qquad \mbox{(Type $S\,I$)}\qquad ,
\end{equation}  
it may correspond to any type $B$ Bianchi space, namely: Bianchi type $V$ if $a=b=0$,   type $IV$ if $a$ or $b$ non-zero, and otherwise types $III$, $VI_h$ for $ab<0$
or $VII_h$ for $ab>0$ with $h=k^2/(a\,b)$.
If it has a double null eigenvector, then :
\begin{equation}\label{structCstSII}
(n_{S\,II}^{\alpha\,\beta})= \left( \begin{array}{ccc}
    \nu+a & \nu & 0 \\ 
   \nu & \nu-a & 0 \\ 
    0 & 0 & 0 \\ 
  \end{array}\right)\quad ,\quad \nu=\pm1 \qquad \qquad \mbox{(Type $S\,II$)}\qquad ;
\end{equation}
and corresponds to Lie algebras of Bianchi type $IV$ if $a=0$, otherwise of Bianchi types $III$ and $VI_h$ with $h=-k^2/a^2$.
If the structure tensor density does not have any other eigenvector, it can be put in the form :
\begin{equation}\label{structCstSIII}
(n_{S\,III}^{\alpha\,\beta})= \left( \begin{array}{ccc}
   0 & \nu & 0 \\ 
   \nu & a & 0 \\ 
    0 & 0 & 0 \\ 
  \end{array}\right)\quad ,\quad 4\,\nu^2>a^2 \qquad \qquad \mbox{(Type $S\,III$)}\qquad 
\end{equation}
and corresponds to Bianchi types $III$ and $VI_h$ with $h=-k^2/\nu^2$.\\

 \item{Lightlike $k_\alpha$ : }  Here, without lost of generality, we may assume $k_\alpha=(1,1,0)$. Using the Jacobi identity we obtain the expression  
\begin{equation}
(n^{\alpha\,\beta})= \left( \begin{array}{ccc}
     a & -a & b \\ 
   -a & a & -b \\ 
    b & -b & c \\ 
  \end{array}\right) 
\end{equation} 
\end{itemize}
corresponding to Bianchi type $V$ if $a=b=c=0$, type $IV$ if $b^2=ac$ and Bianchi types $III$, $VI_h$ for $b^2>ac$ or 
$VII_h$ for $b^2<ac$ with $h=1/(a\,c-b^2)$ in the other cases.
Without lost of generality, it can still be simplified, by performing an appropriate null rotation around $k_a$, that leads to
\begin{equation}
(n_L^{\alpha\,\beta})= \left( \begin{array}{ccc}
     a & -a & 0 \\ 
   -a & a & 0 \\ 
    0 & 0 & c \\ 
  \end{array}\right) \qquad \mbox{(Type $L$)}  \qquad\text{or}   \qquad
  (n_{L'}^{\alpha\,\beta})= \left( \begin{array}{ccc}
     0 & 0 & b \\ 
   0 & 0 & -b \\ 
    b & -b & 0 \\ 
  \end{array}\right) \qquad \mbox{(Type $L'$)}\qquad .
\end{equation}%
\section{Solutions of the field equations}
\label{Hom.Sol}

\subsection{Simply transitive groups}
\label{Sol.trans}

{In this section we solve the MG field equations, which were presented in section \ref{MGeqs}, in terms of homogeneous spaces of section \ref{Hom.geom}.}

As already mentioned, in the framework of Bianchi spaces, the field equations 
reduce to algebraic equations. To solve these equations, we first obtain the link between the cosmological and 
coupling constants $\Lambda$ and $\mu$ ($resp.$ $\xi\,$) and the structure constant parameters. Then we insert them into 
the field equations and discuss the remaining constraints that have to be satisfied.

We shall provide some details in the first case, whereas for the other ones we shall just display the solutions of the equations.
\begin{itemize}
\item{Unimodular Lie Algebras}

\subitem{    \bf Type $I$} 
\subsubitem{    $\star$ TMG: }  
Using the expression \eqn{nI} of the tensor density defining the structure constants, the TMG field equations reduce to three independent equations
\begin{eqnarray}
&&2\,a^2 (a + b) + 8\,b\,c^2 + (a^2 - 4\,c^2 + 4 \Lambda) \mu=0\qquad ,\label{TI1}\\
&&\frac{a^2}4 + a\,b + c^2 - 3 \Lambda=0\qquad ,\label{TI2}\\
&&c (-a^2 + 4 a\,b + 8\,b^2 + 4\,c^2 - 2 (a + 2 b) \mu)=0\qquad .\label{TI3}
\end{eqnarray}
First let us assume $c=0$, so that Eq. \eqn{TI3} is trivially satisfied. We deduce from the other two that:
\begin{eqnarray}
&&\Lambda=\frac 1{12}a(a+4\,b)\qquad,\label{vLI}\\
&&\mu=-\frac 32\,a\qquad,\label{vmuI}
\end{eqnarray}
from which we obtain the  Petrov type $D_t$ solution :
\begin{equation} a=-\frac 23 \mu\qquad,\qquad b= \frac {\mu^2-27\,\Lambda}{6\,\mu}\qquad,\qquad c=0\qquad .\end{equation}
To isolate the value \eqn{vmuI} of $\mu$ in Eq. \eqn{TI1} we have  assumed that $a^2   + 4 \Lambda\neq 0$. 
If  $a^2   + 4 \Lambda = 0$, the parameter $\mu$ remains undetermined and 
the field equations are satisfied if the cosmological constant is negative and

\begin{equation}a=-b=2\sqrt{-\Lambda}\qquad, \end{equation}

or if $\Lambda=0$, for $a=0$ and $b$ arbitrary. Of course these last two solutions correspond to conformally flat  spacetimes (Petrov type $O$). More precisely
if $a=0$ the solution is flat; otherwise it is $AdS_3$ when $a=-b\neq 0$.\\
If we assume $c\neq 0$, the values of $\Lambda$ and $\mu$ provided by the first two field equations \eqn{TI1},\eqn{TI2} become
\begin{eqnarray}
&&\Lambda=\frac 1{12}(a^2+4\,a\,b+4\,c^2)\label{cvLI}\qquad ,\\
&&\mu=-\frac 32\left( \frac{a^2(a+b)+4\,b\,c^2}{a(a+b)-2 \,c^2}\right)\label{cvmuI}\qquad .
\end{eqnarray}
Inserting these values into the third nontrivial field equation \eqn{TI3} we obtain
\begin{equation} ((a + b)^2 - c^2) (a^2 + 4 a\,b + 4\,c^2)=0\qquad,\end{equation}
where we have assumed that $\mu\neq 0$. This leads to two solutions (changing the sign of $c$ correspond to a reflexion of $\theta^1$ or $\theta^2$ and thus to change the sign of $\mu$) of Petrov type $D_s$ :
\begin{equation}a=\frac{27\, \Lambda - \mu^2}{6\, \mu}\qquad,\qquad b= \frac{5\, \mu^2-27\, \Lambda }{12\, \mu}\qquad,\qquad c=\pm(a+b)= \pm\frac{  \mu^2+9\, \Lambda }{4\, \mu}\end{equation}
and a third one, of Petrov type $I_{\mathbb R}$, non flat, but with vanishing cosmological constant, and
\begin{equation} b=\frac 12(\mu+a)\qquad,\qquad c^2= -\frac  14\,a( 3\, a +2\,\mu)\qquad,\end{equation}
which restricts the parameter $a$ to belongs to the interval\footnote{When we write an interval as $[x_0,x_1]$ we 
did not assume $x_0\leqslant x_1$, but  consider the union of the sets $\{x\vert x_0\leqslant x\leqslant x_1\}\cup\{x\vert x_1\leqslant x\leqslant x_0\}$ 
even, unless $x_0=x_1$, one of these two sets is always empty. This convention allows to avoid tedious (but elementary) discussion about signs.}: $a \in\  [0,-\frac 23 \mu]$ .

\subsubitem{ $\star$    NMG: } 
The strategy is the same, but the equations a little bit more cumbersome. The field equations lead to
\begin{eqnarray}
&&\hspace{-12mm}-21a^4-20a^3b+80abc^2+256b^2c^2+80c^4+8\xi\,(a^2+2ab-4\Lambda)=0\qquad , \label{NnI1}\\
  &&\hspace{-12mm}-63\,a^4-80\,a^3\,b+8\,a^2\,\left(-2\,b^2+3\,c^2+2\,\xi\,^2\right)+16\,\left(16\,b^2\,c^2+5\,c^4-4\,c^2\,\xi\,+4\,\Lambda\, \xi
   ^2\right)=0\quad  ,\label{NnI2}\\
   &&\hspace{-12mm}c\,\left(5\,a^3-2\,a^2\,b+40\,a\,b^2-8\,\xi\,\,(a+2\,b)+20\,a\,c^2+64\,b^3+104\,b\,c^2\right)=0\quad . \label{NnI3}
\end{eqnarray}
In this case, there are also  three different types of solutions. \\ 
If we assume $c=0$, we obtain
\begin{eqnarray}
&&\Lambda=\frac{a(21\,a^2+72\,a\,b+16\,b^2)}{8\,(21\,a+4\,b)}\qquad,\label{LnIN}\\
&&\xi\,=\frac a8(21\,a+4\,b)\qquad,\label{XnIN}
\end{eqnarray}
which leads to a  Petrov type $D_t$ solution :
\begin{eqnarray}
&&a^2=\frac{16}{21} \left(3\,\xi\,\pm \sqrt{3}\,\sqrt{\xi\,\,\left(7\,\Lambda +5\,\xi\,\right)} \right)\qquad ,\\
&&b^2=\frac{42\,\Lambda\, \xi\,\pm\sqrt{3}(21\,\Lambda -17\,\xi\,)
   \sqrt{\xi\, \left(7\,\Lambda +5\,\xi\,\right)}-66\,\xi^2}{4(\Lambda -\xi\,)}\qquad .\end{eqnarray} 
We find more convenient to discuss this solution by introducing the parameter $x=b/a$ in terms of which we may rewrite eqs \eqn{LnIN}, \eqn{XnIN} as
\begin{equation}
 \Lambda=\frac{a^2}8\,\frac{21 +72\,x+16\,x^2}{21 +4\,x}\qquad ,\qquad
 \xi\,=\frac {a^2}8\,(21 +4\,x)\qquad .
\end{equation}
In order to have $\xi\,>0$ we need, in addition to $a\neq 0$, to impose that  $x>-21/4$, which implies that $\Lambda>0$ if 
$ -(2\sqrt{15}+9 )/4>x>-21/4$ or $x> (2\sqrt{15}-9 )/4$ .\\
If we did not assume $c=0$, the first two equations \eqn{NnI1}, \eqn{NnI2} lead to much more complicated expressions
\begin{eqnarray}
&&\Lambda=\frac{2\,a c^2 \left(15 a^3+28 a^2 b-128\,b^3\right)+a^3 (a+b) \left(21 a^2+72\,a b+16 b^2\right)-16 b\,c^4 (15 a+32
   b)-160 c^6}{8 \left( a^2 (a+b) (21 a+4 b)-2 c^2 \left(3 a^2+10 a b+64 b^2\right)-40 c^4\right)}\nonumber\\
   && \\
&&\xi\,=\frac{a^2 (a+b) (21 a+4 b)-2 c^2 \left(3 a^2+10 a b+64 b^2\right)-40 c^4}{16\left( a (a+b)-2\, c^2\right)}\qquad,
\end{eqnarray}
inserting these into the field equation \eqn{NnI3} we find
\begin{equation}a (a + b)^2 (a^2 + 2\,a\,b - 4 b^2)-(a^3 + 10 a^2 b + 12\,a b^2 + 8\,b^3)c^2+8\,b\,c^4=0\end{equation}
whose solutions are $c^2=(a+b)^2=a^2(1+x)^2$  and $ c^2=a(a^2+2\,a\,b-4\,b^2)/8\,b=a^2 (1+2\,x-4\,x^2)/{8\,x}$~.\\
From the first solution, which is Petrov type $D_s$, we obtain
\begin{equation}  \xi\,=\frac{a^2}{16} (1+2\,x)(25+42\,x)\qquad,\qquad \Lambda=\frac{a^2}8\frac{(1+2x)(84\,x^2+228\,x+109)}{(25+42\,x)}\qquad,\end{equation}
which satisfy the requirement $\xi\,>0$ if $x\not\in [-25/42,-1/2]$. 
We can immediately see that this restriction is compatible with both signs of the cosmological constant : 
$\Lambda <0$ if  $x  \in ]-(8 \sqrt{15}+57)/42,\,(8 \sqrt{15}-57)/42[\,\cup\,]-25/42,-1/2[$ ; $\Lambda=0$ at the boundary of these intervals, excepted at $x=-25/42$ where it diverges ; 
otherwise $\Lambda>0$  if $x  \not\in [-(8 \sqrt{15}+57)/42,\,(8 \sqrt{15}-57)/42]\,\cup\,]-25/42,-1/2]$. 
Thus  $\Lambda$ is bounded from below (by approximatively $-0.3\, a^2$), but it can be chosen arbitrarily positive.

For the second solution $ c^2=a^2 (1+2\,x-4\,x^2)/{8\,x}$ things are a little bit more involved. We obtain  
\begin{equation} 
\xi\,=\frac{a^2}{32\,x} (64\,x^3-44\,x^2+36\,x+5)\qquad,\qquad \Lambda=\frac{5\,a^2}{16\,x}\frac{(1+2x)^4}{(64\,x^3-44\,x^2+36\,x+5)}\qquad .
\end{equation}
Thus $\xi\,$ will be positive only if $x\not \in [x_0,\,0]$  where $x_0\simeq-0.119$ is the single real root of the cubic polynomial 
$64\,x^3-44\,x^2+36\,x+5$. For $x\in [x_0,\,0]$ we always have $\Lambda<0$ ; for $x\not \in [x_0,\,0]$ we have $\Lambda>0 $, a decreasing 
function of $x$ which varies from $+\infty$ to $5\,a^2/64$ .

\subsubitem{ $\star$    GMG:} 
The combination of the two theories,  introduces three constants that are related to the components of the structure tensor density \eqn{nI} by
\begin{eqnarray}
&&\Lambda=-\frac{5 \left(a^2+4\, a\, b+4\,c^2\right)^3}{8 \left(a^2(3\, a^2-40\, a\, b-112\, b^2)-8(9\,a^2+20\, a\,b\,-16\,b^2) c^2-80\,
   c^4\right)}\label{LaGMGI}\\
   &&\mu=\frac{a^2(-3\, a^2+40\, a\, b+112\, b^2)+8(9\,a^2+20\, a\,b\,-16\,b^2) c^2+80\, c^4}{16 \left(a^3+2\,a^2 b-4 \,a\, b^2-8\,b
   c^2\right)}\label{MuGMGI}\\
   &&\xi\,=\frac{ a^2(-3\, a^2 + 40\, a\, b + 112\, b^2) +8(9\,a^2+20\, a\,b\,-16\,b^2) c^2+80 c^4 }{ 8 (a^2 + 4\, a\,b + 4\,c^2) }\label{XiGMGI}
\end{eqnarray}
Inverting this system as well as  discussing in general the positivity of $\xi\,$ is not very illuminating. 
So, we have plotted the region in the ($x=b/a$ , $y=c/a$) plane of the sign of $\xi\,$ on Fig. \ref{figLXMnI}. 
{\begin{figure}[h]
\begin{center}
\includegraphics[height=75mm]{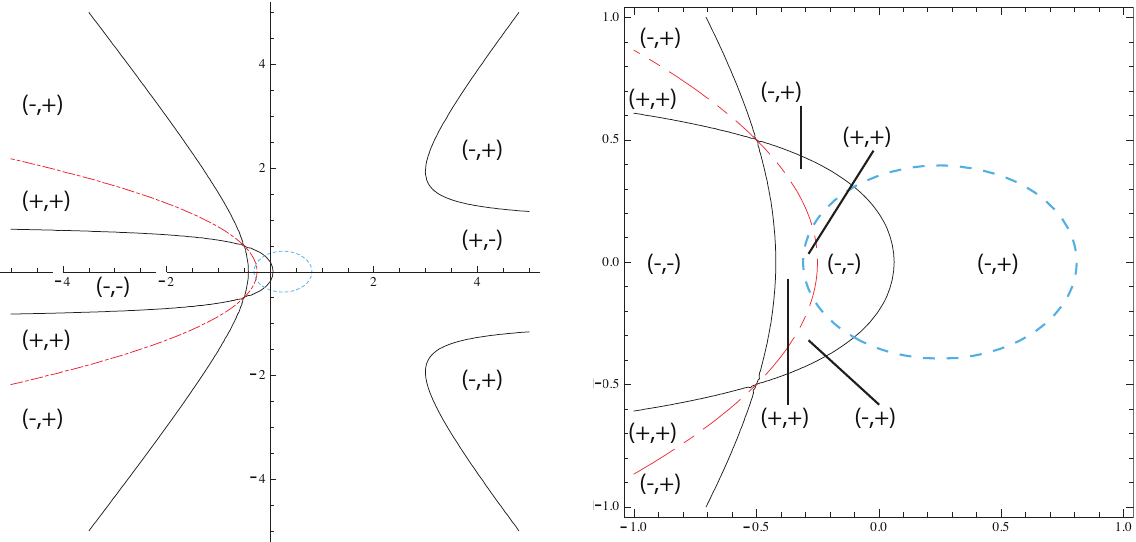}
\end{center}
\caption{\footnotesize Graphical representation of the sign of the GMG coupling constant $\Lambda$ and $\mu$ as function of 
the rescaled parameters $x=b/a$ and $y=c/a$   in the framework of unimodular group spaces of type $I$. We have $\Lambda=\infty$, 
$\xi\,=0$, $\mu=0$ on the solid curve; $\xi\,=\infty$, $\Lambda=0$ on the dash-dotted curve; $\mu=\infty$ on the dotted curve. 
On each region of the $x$, $y$ plane delimited by these curves we have indicated the sign of $\Lambda$ (equal to the one of $\xi\,$) 
and the sign of $\mu$. The left--hand side of the figure is a blow--up of the central part of the right--hand side plot.}
\label{figLXMnI}
\end{figure}}
Let us also remark that: $\Lambda \,\xi\,=(5/64) (a^2+4\,a\,b+4\,c^2)^2  \geqslant 0$.\\ 
However, using the special values of the structure tensor density we can easily see from condition \eqn{stabcond} that there are  Petrov
type $O$ solutions, with arbitrary value of $\mu$ and negative values of $\Lambda=-(45/184)\,a^2$ and $\xi=-(23/8)\,a^2$), that are stable for $\mu^2\leqslant(12167/16208)\,a^2$. There are also Petrov types $D_t$ and $D_s$ solutions that allow positive values of $\xi$ . As for a structure 
tensor density of type $I$, we have as many independent\footnote{ Generically, the Jacobian ($J$) of the transformation defined by the equations (\ref{LaGMGI}-\ref{XiGMGI}) is: $ J=40\, a\,c\,\frac{((a+b)^2-c^2)(c^2-b^2)(a^2+4\,a\,b+4\,c^2)}{(a^3+2\,a^2\,b-4\,a\,b^2-8\,b\,c^2)^3}$ .}  parameters in the solution as there are physical constants of the problem. Thus 
we may expect that for some range of values a finite number of solutions are always defined. Of course, when the number 
of geometrical parameters will be less than three, the solutions, if any, will exist only for special values of the physical constants.

\subitem{ \bf Type $II$} 

\subsubitem{ $\star$  TMG: }  Three different solutions occur. One, of Petrov type $N$, with negative cosmological constant
\begin{equation}\Lambda=-\frac {\mu^2}9\qquad,\qquad a=-\frac {2\,\mu} 3\qquad,\qquad b=\frac {2\,\mu} 3 \qquad;\end{equation}
and two solutions, with vanishing cosmological constant, of Petrov types $N$  and $II$ respectively, 
\begin{equation}\label{II.1}\Lambda=0\qquad,\qquad a=-\frac \mu 2\qquad,\qquad b=0\qquad;\end{equation}
\begin{equation}\label{II.2}\Lambda=0\qquad,\qquad a=-\frac {\mu} 6\qquad,\qquad b=\frac {2\,\mu} 3\qquad.\end{equation}
Let us mention that trivial flat solutions also occur as solutions of the field equations, for 
instance we find a solution with $a=b=0$, corresponding to the flat space\footnote{This emphasises the fact that Bianchi $I$ is always a solution, it simply corresponds to a flat space, 
but flat space may also appear as a Bianchi $II$ model. In the same way anti-de Sitter space can be seen as a Bianchi $VIII$ space but, locally, also as Bianchi $III$ model.} ; we shall not insist anymore on such solutions.
\subsubitem{ $\star$  NMG: }  We also obtain  three types of solutions; two of Petrov type $N$:
\begin{equation}b=0\qquad,\qquad a^2=\frac 14\,\xi\,\qquad,\qquad\Lambda=0 \end{equation}
\begin{equation}b=-a\qquad,\qquad a^2=\frac 8{17}\,\xi\,\qquad,\qquad\Lambda=-\frac{35}{289}\,\xi\, \end{equation}
and a third one of Petrov type $II$ :  
\begin{equation}\label{II.3}b=-(1\pm\sqrt 5)a\qquad,\qquad  a^2=\frac {(61\mp19\sqrt{5})}{479}\,\xi\,\qquad,\qquad
\Lambda= \frac{5(3521\mp1560\,\sqrt{5})}{229441}\,\xi\,\qquad . \end{equation}

Let us notice that the cosmological constant as well as the coupling constant are independent of $\nu$ (see Eq. (\ref{tensor.II}) ), but  the Riemann curvature tensor is $\nu$ dependent.
\subsubitem{ $\star$  GMG: } We obtain solutions of Petrov type $II$:
\begin{eqnarray}
\label{II.4}
&&\Lambda=-\frac{5 b (4 a+b)^3}{8 \left(-112\,a^2-40 a b+3 b^2\right)}\qquad,\\
   &&\mu=\frac{b \left(-112\,a^2-40 a b+3 b^2\right)}{16 \left(-4 a^2+2\,a b+b^2\right)}\qquad,\nonumber\\
   &&\xi\,=\frac{b \left(112\,a^2+40 a b-3 b^2\right)}{8 (4 a+b)}\nonumber\qquad,
   \end{eqnarray}
   where $\Lambda\,\xi\,=(5/32)b^2(4 \,a +b)^2\geqslant 0$. To have $\xi\,>0$,  $b$ must be in 
   the intervals $ ]a\,(20+4\sqrt{46})/3,\,0[\cup\,]a\,(20-4\sqrt{46})/3,\,-4\,a[$. 
   \\ Solutions of Petrov type $N$ also occur in the cases where $b=0$
 \begin{eqnarray}
\Lambda=0\qquad,\qquad
 \xi\,=\frac{4\,a^2\,\mu}{2\, a+\mu}\qquad,
 \end{eqnarray}
  or when $b=-a$
 \begin{eqnarray}
\Lambda=-a^2\frac{70\, \mu+3\,a}{272\,\mu}\qquad,\qquad
 \xi\,=\frac{17\,a^2\,\mu}{ 4(3\, a+2\,\mu)}\qquad.\end{eqnarray}

   \subitem{ \bf Type $III$}  

\subsubitem{ $\star$  TMG: } The condition $\mu\neq 0$ is incompatible with the field equations.

\subsubitem{ $\star$  NMG: }  There is no  solution with $\xi\,\neq 0$.
\subsubitem{ $\star$ GMG: } There is no solution with $\xi\,>0$ but a special one\footnote{We shall make it more explicit in section \ref{coord.repr.SII}.}, of Petrov type $III$ with $\xi\,\leqslant0$, namely :
\begin{equation}
\label{III.GMG}
\mu=-\frac{69}{80}\,a\qquad,\qquad \Lambda=-\frac{45}{184}\,a^2\leqslant0\qquad,\qquad \xi\,=-\frac{23}{8}\,a^2\leqslant0\qquad.
\end{equation}

\subitem{ \bf Type $IV$}
\subsubitem{ $\star$  TMG: }  Taking into account the condition $4\,\nu^2 > a^2$, we obtain as the only (real) solution
\begin{equation}a=\frac \mu 2\mp \nu\qquad,\qquad b=\frac{\mu}2 \pm  \nu\qquad,\qquad\Lambda=0\end{equation}
of Petrov type $I_{\mathbb C}$, and subject to the condition : $\nu\not \in [\mp\mu/2,\,\pm\mu/6]$ .
\subsubitem{ $\star$  NMG: }   Once $\Lambda$ and $\xi\,$ are expressed in terms of $a$, $b$ and $\nu$, the field equations are satisfied if :
\begin{equation}\nu^2=b(a-b)\qquad\text{or}\qquad \nu^2=\frac{(a^2-b^2)(a-b)}{4\,a}\qquad .\end{equation}
It is easy to verify that the first one is disallowed by the condition $4\,\nu^2>a^2$. 
Whereas for the second one, we may parametrise again the solutions as follows :
\begin{eqnarray}
&&b=x\,a\qquad ,\\
&&\nu^2=\frac {a^2}4\,(1-x^2)(1+x)\qquad ,\\
&&\Lambda=\frac {5\,a^2}8\,\frac{(1-x)^4x^2}{(8+11\,x+18\,x^2-5\,x^3)}\qquad ,\\
&&\xi\,=\frac {a^2}8\,(8+11\,x+18\,x^2-5\,x^3)\qquad .
\end{eqnarray}
The condition $4\,\nu^2>a^2$ implies that $x>(1+\sqrt{5})/2$ or $0>x>(1-\sqrt{5})/2$ while the positivity of $\xi\,$ 
requires $x<x_0$ where $x_0$ is the single real root of the cubic polynomial $8+11\,x+18\,x^2-5\,x^3$, {\it i.e.} 
$x_0=\small{ \left(18+\sqrt[3]{12987-60 \sqrt{14370}}+\sqrt[3]{3 (4329+20 \sqrt{14370} )} \right)/15}$,
 $x_0\approx 4.21$, and insures that $\Lambda >0$. This solution is of  Petrov type $I_{\mathbb C}$, both 
 with respect to the classification of the Ricci and the Cotton--York tensors.
\subsubitem{ $\star$  GMG: } The generic solution is of Petrov type $I_{\mathbb C}$ with
 \begin{eqnarray}
&&\Lambda=-\frac{5 \left[(a-b)^2-4 \nu ^2\right]^3}{8 \left[(a-b)^2 \left(3 a^2+26 a b+3 b^2\right)+8 \nu ^2 (a-9 b) (a-b)-80
   \nu ^4\right]}\qquad ,\\
&&\mu=\frac{(a-b)^2 \left(3 a^2+26 a b+3 b^2\right)+8 \nu ^2 (a-9 b) (a-b)-80 \nu ^4}{16 \left[(a-b)^2 (a+b)- 4\, a \nu ^2\right]}\qquad ,\\
   &&\xi\,=-\frac{(a-b)^2 \left(3 a^2+26 a b+3 b^2\right)+8 \nu ^2 (a-9 b) (a-b)-80 \nu ^4}{8 \left[(a-b)^2-4 \nu ^2\right]}\qquad ,
   \end{eqnarray} 
where $\Lambda\,\xi\,=(5/64)((a-b)^2-4\nu^2)^2\geqslant 0$. We have plotted on Fig. \ref{figLXMnIV} the zero and singular curves of  
$\Lambda$, $\mu$ and $\xi\,$ on the ($x=a/\nu$, $y=b/\nu$) plane.\par\noindent
We also found a Petrov type $D_s$ solution :
\begin{equation}
a=b=-\frac{\xi\,}{2\mu}~\qquad,\qquad \nu^2=-\frac 25\,\xi\,>0\qquad,\qquad  {\rm with:}~\Lambda=\frac 15\,\xi\qquad.
\end{equation}
{\begin{figure}[h]
\begin{center}
\includegraphics[height=60mm]{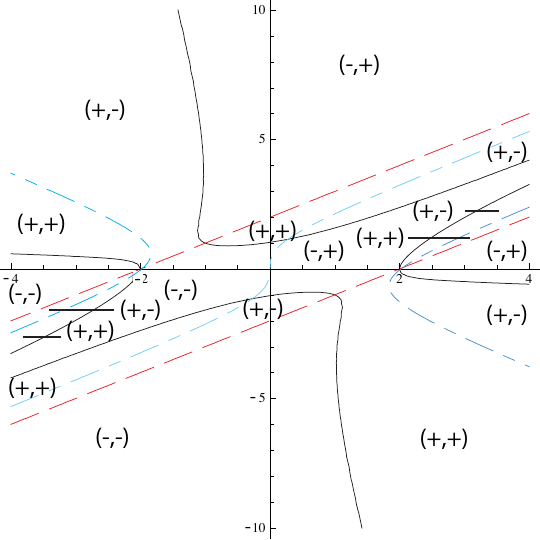}
\end{center}
\caption{\footnotesize Graphical representation of the sign of the GMG coupling constant $\Lambda$ and $\mu$ as function 
of the rescaled parameters $x=a/\nu$ and $y=b/\nu$   in the framework of   unimodular group spaces of type $IV$. 
We have $\Lambda=\infty$, $\xi\,=0$, $\mu=0$ on the solid curve; $\xi\,=\infty$, $\Lambda=0$ on the  two dotted 
straight lines; $\mu=\infty$ on the dash-dotted curve. 
On each region of the $x$, $y$ plane delimited by these 
curves we have indicated the sign of $\Lambda$ (equal to the one of $\xi\,$) and the sign of $\mu$. }
\label{figLXMnIV}
\end{figure}}
\item{Non-unimodular Lie Algebras}
\subitem{ \bf Type $T$}
\subsubitem{ $\star$  TMG: }  
We obtained two solutions, only defined for positive cosmological constant. The first one is of Petrov $O$ 
and locally a $dS_3$ space, 
\begin{equation}  b=0\qquad,\qquad k^2=\Lambda\qquad ,\end{equation}
while the second one is of Petrov Type $D_s$
\begin{equation}
a=\frac \mu 3 \qquad,\qquad k^2=\frac 34\,\Lambda-\frac 1{36}\, \mu^2\qquad,\qquad b^2=\frac 34\,\Lambda+\frac 1{12}\,\mu^2~,\qquad \Lambda> \mu^2/27~.
\end{equation}

\subsubitem{ $\star$  NMG: }  Three different solutions are available.
The first one is of Petrov type $I_{\mathbb R}$ (or Petrov type $D_s$ when $\xi\,=2\,\Lambda$)
\begin{equation}a=0\qquad,\qquad b^2=\frac 1{8}(3\,\xi\,+ \Lambda)\qquad,\qquad k^2=\frac 1{8}(5\,\Lambda-\xi\,)\end{equation} 
which of course requires that $\Lambda >\xi\,/5$ .\\
The second one is of Petrov type $O$ and locally a $dS_3$ space :
\begin{equation}
b=0\qquad,\qquad  k^2=2\left(\xi\, \pm\sqrt{\xi\,(\xi\,- \Lambda})\right)\qquad.
\end{equation} 
Note that $\xi\notin]0,\Lambda[$; and $k^2\geqslant 0$ requires the plus sign in front of the 
square root  unless if $\xi\geqslant \Lambda\geqslant0$ in which case both signs are accepted.

The third  one, of Petrov type $D_s$ :
\begin{equation}
\label{TypeT.Ds.NMG} 
b^2=a^2+k^2\qquad,\qquad a^2=\frac{6\,\xi\,\mp\sqrt{\xi\,(15\,\xi\,+21\,\Lambda)}}{21}
\qquad,\qquad k^2=\frac{\sqrt{\xi\,(15\,\xi\,+21\,\Lambda)}\pm4\,\xi\,}{4}
\end{equation}
which are real for the upper signs if $\xi>0$ and for the lower ones if $\xi\in[0,21\Lambda]$ or 
$\xi\in[\Lambda,0]$ depending if $\Lambda$ is positive or negative respectively.
 Note that for $\xi=\Lambda$ the Cotton--York tensor is vanishing while the traceless Ricci 
and $\widehat K_{\alpha\,\beta}$ tensors remain of Petrov Type $D_s$.\\

\subsubitem{ $\star$  GMG: } Generically we obtain solutions of Petrov type $I_{\mathbb R}$ :
\begin{eqnarray}
&&\Lambda=\frac{3\,k^4-14\,b^2\,k^2+16\,a^2\,k^2-5\,b^4}{2(k^2+8\,a^2-5\,b^2)}\qquad,\\
&&\mu=\frac{k^2+8\,a^2-5\,b^2}{8\,a}\qquad,\\
&&\xi\,=\frac{-k^2-8\,a^2+5\,b^2}2\qquad.
\end{eqnarray} 
The singular and zero curves of $\Lambda$, $\mu$ and $\xi\,$ in the ($x=a/k$, $y=b/k$) plane are depicted on Fig.~\ref{figLXMnT}. Let us mention that in the region where $\xi\,>0$ (the interior of the hyperbola) we have also that $\Lambda >0$. \\
{\begin{figure}[h]
\begin{center}
\includegraphics[height=60mm]{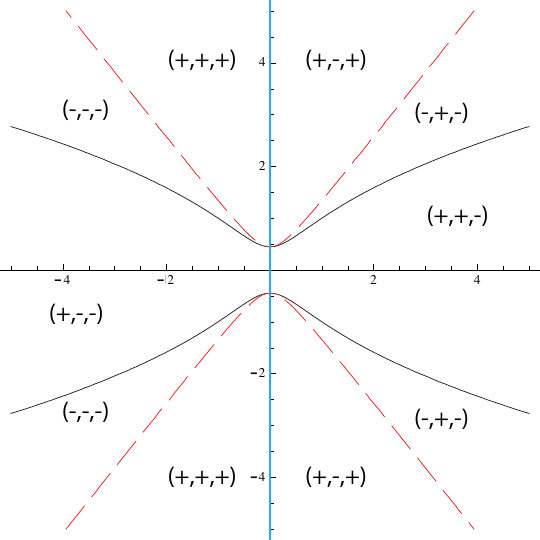}
\end{center}
\caption{\footnotesize Graphical representation, in the framework of non-unimodular group spaces of type $T$, of the zero and 
singular curves of the cosmological constant $\Lambda$ and of the  coupling constants $\xi\,$ and $\mu$ as function of the 
rescaled parameters $x=a/k$ and $y=b/k$ ( $\Lambda=\infty$ but $\xi\,=0$ and $\mu=0$ on the (dashed) hyperbola, $\Lambda=0$ 
on the black fourth order algebraic (solid) curve, $\mu=\infty$ on the $y$ axis ). For each region of the plane delimited 
by these curves, we indicate the sign of $\Lambda$, $\mu$ and $\xi$.}
\label{figLXMnT}
\end{figure}}
There are also solutions of Petrov type $D_s$, with $k^2=b^2-a^2$, $\mu$ arbitrary  and
\begin{equation}
\label{TypeT.Ds.GMG} 
\Lambda=\frac{3 a^5-35 a^4 \mu -8 a^3 b^2+40 a^2 b^2 \mu -16 a b^4+16 b^4 \mu }{2 \mu 
   \left(17 a^2+4 b^2\right)}\qquad,\qquad
 \xi\,=\mu\frac{17\, a^2 + 4 \,b^2 }{2(\mu -3\,a)}\,.
\end{equation}
We plot on Fig. \ref{figLXMnTDs} the zero curve of $\Lambda$ and the singular straight line of $\xi\,$ in the ($x=a/\mu$, $y=b/\mu$) plane. 
{\begin{figure}[h]
\begin{center}
\includegraphics[height=60mm]{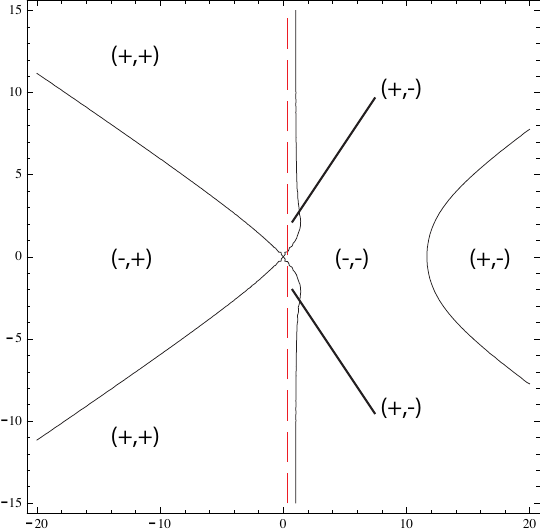}
\end{center}
\caption{\footnotesize Graphical representation of the singular line ($x=1/3$) of $\xi\,$ and the zero  curve of the 
cosmological constant $\Lambda$   of the Petrov type $D_s$ solution of $GMG$ field equations, in the framework of 
non-unimodular group spaces of type $T$. The zero curve of $\Lambda$ is a fifth order algebraic curve, admitting 
the vertical asymptote $x=1$ and the two oblique asymptotes $y=\pm x/2$. }
\label{figLXMnTDs}
\end{figure}}

Finally, solutions of Petrov type $O$ are obtained for arbitrary values of the coupling constant $\mu$ and $\xi$. These are $dS_3$ (or flat) spaces but with
\begin{equation}
\Lambda=\frac{4\,\xi\,\,-k^2}{4\,\xi\,}~k^2~.
\end{equation}   
Let us notice that for $a=0$,  the solutions \eqn{TypeT.Ds.NMG}  and \eqn{TypeT.Ds.GMG} correspond to conformally flat geometries with $\Lambda=\xi\gtrless0$, like $\mathbb{R}\times dS_2$ or $\mathbb{R}\times AdS_2$ respectively, the latter having been considered by Cl\'ement  \cite{NMGSol}.
\subitem{ \bf Type $S\,I$} 
\subsubitem{ $\star$  TMG: }  Two non nontrivial solutions occur. The first one, is Petrov type $O$ and locally an $AdS_3$ space with $\Lambda<0$ ,
 \begin{equation}b=-a \qquad,\qquad k^2=-\Lambda\qquad .\end{equation}
The second one, with $\mu^2/27> \Lambda \geqslant -\mu^2/9$ , is of Petrov type $D$ :
\begin{equation}a=\frac {-2\mu\pm \sqrt{27\,\Lambda+3\,\mu^2}}6\qquad,\qquad b=\frac {2\mu\pm \sqrt{27\,\Lambda+3\,\mu^2}}6\qquad,\qquad k^2=-a\,b=  \frac{\mu^2-27\,\Lambda}{36}\qquad .\end{equation}
More precisely, with the upper sign $(+)$, the solution is of Petrov type $D_s$ ($resp.$ $D_t$) for $\mu>0$ ($resp.$ $\mu<0$) ; with the lower sign $(-)$, 
it is the converse : $D_t$ ($resp.$ $D_s$) for $\mu>0$ ($resp.$ $\mu<0$).\\

\subsubitem{ $\star$  NMG: }  
Here  we obtain $\xi\,=\frac 18[5\,(a+b)^2+16\,(a^2+b^2)+36\,k^2]>0$ and the remaining field equations are satisfied in three cases. If :
\begin{equation}a=-b\qquad,\qquad k^2={2\left(-\xi\,\pm\sqrt{\xi\,(\xi\,- \Lambda)}\right)}\qquad;\end{equation}
 the minus sign requires that $0>\Lambda\geqslant \xi\,$, whereas for the plus sign : 
$\xi\,>0$ and $\Lambda<0$ or $\xi\,<0$ and $\Lambda\geqslant\xi\,$. It is a Petrov type $O$ solution, corresponding to an $AdS_3$ space, which is stable when $\xi>0$ .\\
We also obtain  
$k^2=-a\,b$, in which case, parametrising the solution as before by  $b=x\,a$, we have
\begin{equation}\xi\,=\frac {a^2}8[8(x^2+1)+13(x-1)^2]>0\qquad\text{and}\qquad \Lambda=\frac {a^4} {64\,\xi\,}(21\,x^4+204\,x^3-194\,x^2+204\,x+21)\qquad .\end{equation}
The positivity of $k^2$ requires that $x<0$. So the maximum of $\Lambda/ \xi\,$ is reached at $x=0$ or $x=-\infty$ where the ratio tends to $1/21$ . 
In the limit $x=0$, we obtain $\Lambda= \xi\,/21=a^2/8$. As the quartic polynomial  $21\,x^4+204\,x^3-194\,x^2+204\,x+21$ has only two (negative) 
real roots : $x_1\approx -0.1$ and $x_2\approx -10.7$, $\Lambda$ is negative for the value of $x$ between these two roots. It reaches its minimum 
at $x\approx -1$. The coupling constant $\xi\,$ increases monotonically as $x$ decreases;    when $x$ goes to $-\infty$, where both $\Lambda$ 
and $\xi\,$ diverge. This solution is of Petrov type $D_t$ if $x\in ]-1,\,0[$ , Petrov type $O$ when $x=-1$ , and Petrov type $D_s$ if $x\in ]-\infty,\,-1[$ .\\
There  also is a third solution of Petrov type $I_{\mathbb C}$ (unless $a=b=0$)
\begin{eqnarray}
a=b=\pm\frac{\sqrt{\Lambda+3\xi\,/2}}{2\sqrt{2}}~,\qquad k^2=\frac{\xi\,-10\Lambda}{16}\quad , 
\end{eqnarray}
which according to the sign of $\Lambda$  requires $\xi\,>10\Lambda\geqslant0$ or $\xi\,>-\frac{2}{3}\Lambda\geqslant0$ 
to be defined.
\subsubitem{ $\star$  GMG: }

Generically we obtain a Petrov type I solution
\begin{eqnarray}
&&\Lambda=\frac{-5 (a +b )^4-8 \left(a ^2-30 a  b +b ^2\right) k ^2+48 k ^4}
{8 \left(3 a ^2-26 a  b +3 b ^2-4 k ^2\right)}\qquad,\\
&&\mu=\frac{-3 a ^2+26 a  b -3 b ^2+4 k ^2}{16 (a -b )}\qquad,\\
&&\xi\,=\frac{1}{8}\left(-3 a ^2+26 a  b -3 b ^2+4 k ^2\right)\qquad.
\end{eqnarray}
We also recover a Petrov type $O$ solution, namely an $AdS_3$ space, when
\begin{eqnarray}
b=-a\qquad,\qquad \Lambda=-k^2\frac{k^2+4\,\xi\,}{4\,\xi\,}\qquad.
\end{eqnarray}
The absence of the tachyonic massive mode \eqn{stabcond} reads 
\begin{equation}\label{SIstab}
\xi\left(2\xi^2(\xi+4\mu^2)+k^2\mu^2(4\xi+k^2)\right)\geqslant0\qquad .
\end{equation}
Accordingly, at least for large positive or negative values of $\xi$ the solutions will not contain tachyons.

We also obtain special solutions for $k^2=-ab$, in which case we parametrise the solutions as before
by $b=x\, a$ with $x<0$ :
\begin{eqnarray}
&&\Lambda=-a^2\,\frac{\left(-21-55 x+14 x^2-14 x^3+55 x^4+21 x^5\right) \alpha -2 \left(21+204 x-194 x^2+204 x^3+21 x^4\right) \mu}
{16 \left(21(1+x^2)-26 x\right) \mu },\nonumber\\
&&\xi\,= \frac{a^2 \mu}{4}\,\frac{21(1+x^2)-26 x}{3(1-x)a +2\,\mu} \qquad.
\end{eqnarray}
These solutions are of Petrov type $D_t$ if $x\in ]-1,\,0[$, Petrov type $O$ when $x=-1$, and Petrov type $D_s$ if $x\in ]-\infty,\,-1[$.\\

\subitem{ \bf Type $S\,II$} 
\subsubitem{ $\star$  TMG: } We easily obtain a Petrov type $N$ solution
\begin{equation}
\label{SIITMG}
k=\pm\sqrt{ -\Lambda} \qquad,\qquad a=\frac{\mp\sqrt{-\Lambda}-\mu}{2}\qquad ;
\end{equation}
we also recover a Petrov type $O$ solution, namely an $AdS_3$ space when 
\begin{equation}
a=-k\qquad,\qquad\Lambda=-k^2\qquad.
\end{equation}
\subsubitem{ $\star$  NMG: }   
Here, we obtain a Petrov type $N$ solution. If 
\begin{equation}\label{SIINMG}
\Lambda=-\frac{k^2 (4 a+3 k) (4 a+k)}{2 \left(8 a^2+8 a k+k^2\right)}~,\qquad\text{and}\qquad \xi\,=\frac 12\left(8 a^2+8 a k+k^2\right)
\end{equation} 

   the field equations are satisfied. Accordingly, imposing $\xi\,>0$, we obtain solutions for all values of $a$ and $k$ such that
   $k\not \in [-4\,a-2\sqrt{2}\vert a\vert,\, -4\,a+2\sqrt{2}\vert a\vert]$. All these solutions have negative cosmological constant.
   Let us also notice that, here again, the value of the parameter $\nu$ did not play any r\^ole in the parametrisation of the solutions, but only appears in the curvature. 
   For $a=-k$, we reobtain a Petrov type $O$ solution, a stable $AdS_3$ geometry. For $a=-k/2$ we have also a conformally flat geometry,  solving   
   the field equations, but  for negative value of $\xi\,=-2\,a^2$.
   
   \subsubitem{ $\star$  GMG: }
   
Here, we obtain a Petrov type  $N$ solution 
\begin{eqnarray}\label{SIIGMG}
\Lambda=-\frac{16 a ^2 \mu +k ^2 (k +3 \mu )+2 a  k  (k +8 \mu )}{2 \left(8 a ^2
+8 a  k +k ^2\right) \mu }k^2,\qq
\xi\,=\mu \,\frac{8\, a ^2+8\, a\,  k +k ^2}{2(2\, a +k +\mu)}~.
\end{eqnarray}
Moreover, there also exists a solution of Petrov $O$ type for $a=-k$, which is locally an $AdS_3$ space with $\Lambda=-\frac{k^2+4\,\xi\, }{4\,\xi\, }k^2$,  and whose absence of tachyonic mode is also provided by Eq. \eqn{SIstab}.

\subitem{ \bf Type $S\,III$} 
\subsubitem{ $\star$  TMG: }  We  obtain a solution of Petrov type $D$ :
\begin{equation}a=\frac{2\,\mu}3\qquad,\qquad k^2=\nu^2=\frac {\mu^2-27\,\Lambda}{36}\qquad,\qquad \mu^2<-9\Lambda\qquad,\end{equation}
which is of Petrov type $D_s$ ($resp.$ $D_t$) for $k=\nu$ ($resp.$ $k=-\nu$).

\subsubitem{ $\star$  NMG: }  The following set of solutions occurs:

Firstly a  Petrov type $I_{\mathbb R}$ solution :
\begin{equation}a=0\qquad,\qquad k^2=\frac{\xi\,-5\,\Lambda}{8}\qquad,\qquad \nu^2=-\frac{3\,\xi\,+\Lambda}{8}\qquad,\end{equation}
which implies that $\Lambda<-3\,\xi\,$ . 

Secondly, we have also   solutions of Petrov type $D$ :
\begin{equation}k=\pm\nu\quad,\quad a^2= \frac 4{21}\,\left(6\,\xi\,+\sqrt{3\,\xi\,(5\,\xi\,+7\,\Lambda)}\right) \quad,\quad \nu^2=\frac 1{4}\,\left(4\,\xi\,+\sqrt{3\,\xi\,(5\,\xi\,+7\,\Lambda)}\right)\end{equation}
which requires, taking into account the condition $4\,\nu^2>a^2$ and assuming $\xi\,>0$ that 
$\Lambda\geqslant -5\,\xi\,/7$, otherwise $2\,\Lambda\leqslant\xi<0$. \\
Finally, we also obtain :
\begin{equation}\label{KSHNMG}
k=\pm \nu\quad,\quad a^2= \frac 4{21}\,\left(6\,\xi\,-\sqrt{3\,\xi\,(5\,\xi\,+7\,\Lambda)}\right) \quad,\quad 
\nu^2=\frac 1{4}\,\left(4\,\xi\,-\sqrt{3\,\xi\,(5\,\xi\,+7\,\Lambda)}\right)
\end{equation}
which requires  that $0>-35\,\xi\,/289>\Lambda\geqslant -5\,\xi\,/7$ . These solutions are of Petrov type $D_s$ ($resp.$ $D_t$) for $k=\nu$ ($resp.$ $k=-\nu$).

\subsubitem{ $\star$  GMG: } 

Here, we obtain solutions of Petrov type $I$ 
\begin{eqnarray}
&&\Lambda=\frac{-5 a ^4-8 a ^2 \left(5 \nu ^2-k^2\right)+16 \left(3 k ^4-14 k ^2 \nu ^2-5 \nu ^4\right)}
{8 \left(3 a ^2-4 k ^2+20 \nu ^2\right)}\qquad,\\
&&\mu=\frac{3 a ^2-4 k ^2+20 \nu ^2}{16 a }\qquad,\\
&&\xi\,=\frac{1}{8}\left(-3 a ^2+4 \left(k ^2-5 \nu ^2\right)\right)\qquad.
\end{eqnarray}
We also find solutions, with $k=\pm\nu$, of Petrov type $D$
\begin{eqnarray}\label{KSH1GMG}
&&\Lambda=-\frac{21 a ^5-42 a ^4\, \mu -160 a ^3\, \nu ^2+576 a ^2\, \mu\,  \nu ^2+256 a\,  \nu ^4-512 \mu\,  \nu ^4}
{336 \mu\,a ^2 -256 \mu\,  \nu ^2}\qquad,\\
&&\xi\,=\frac{-21 a ^2 \mu +16 \mu\,  \nu ^2}{4(3\,a -2\,\mu)}\qquad,\label{KSH2GMG}
\end{eqnarray}
which are of Petrov type $D_s$ ($resp.$ $D_t$) for $k=\nu$ ($resp.$ $k=-\nu$).

\subitem{ \bf Type $L$} : 
\subsubitem{ $\star$  TMG: }  There is one solution of Petrov Type $D_s$  ,  defined only for special values of $\mu$ and $\Lambda$ (positive):
\begin{equation}\label{L.1} a=-\frac 1c\qquad,\qquad \mu=\pm3\sqrt{3\,\Lambda}\qquad,\qquad c =\pm2\sqrt{3\,\Lambda}\qquad.\end{equation}
\subsubitem{ $\star$  NMG: }   Here also, only one solution of Petrov Type $D_s$   is obtained,  for special values of $\xi\,$ and $\Lambda$ (positive):
\begin{equation}\label{L.2}a = -\frac1c\qquad,\qquad \xi\,=21\,\Lambda\qquad,\qquad c=\pm2\sqrt{2\,\Lambda}\qquad .\end{equation}
\subsubitem{ $\star$  GMG: } There is a solution of Petrov Type $D_s$:  
\begin{equation}\label{L.3}a = -\frac1c\qquad,\qquad \xi\,=\frac{21\mu}{a\,(12+8a\,\mu)}\qquad,\qquad \Lambda=\frac{1+2a\,\mu}{16a^3\mu}\qquad .\end{equation}
We also found a class of solutions for arbitrary $a$ which are of Petrov type $II$:
\begin{equation}
\label{L.4}
\Lambda=-\frac{5}{24}\,c^2\qquad,\qquad \mu=\frac{3  }{16}\,c\qquad,\qquad \xi\,=-\frac{3 }{8}\,c^2\qquad.
\end{equation}

\subitem{ \bf Type $L'$} : The solutions we obtain are flat spaces.
\subsubitem{ $\star$  TMG, NMG and GMG: }  The  only solution is the flat space, obtained with
\begin{equation}
b=0\qquad\text{or}\qquad b=-1\qquad\text{and}\qquad \Lambda=0\qquad.
\label{KLPT}
\end{equation}

\end{itemize}

\subsection{Coordinate representations of the metrics}
\label{coord.repr.SII}

Having at our disposal all the homogeneous solutions of MG theories in a formal way, the purpose of this subsection
is to illustrate how to express them in terms of the invariant forms that define the coordinate system. \par\noindent
This can be achieved in two ways. Either algebraically by determining the $GL(3,\mathbb{R})$ matrix which 
transforms the canonical expression of the structure vector and tensor density in the forms we use or by a direct integration of the Cartan equations \eqn{Cartan} defining the invariant vector 
fields from which we deduce the expression of their dual basis.\\ 
For illustrative purpose we now sketch both approaches.  The first one is performed 
in the framework of the GMG solution \eqn{III.GMG} obtained from the unimodular $Type\ {III}$ structure tensor density\footnote{Assuming $a\neq 0$.} 
\eqn{nIII}; the second one for the non--unimodular $S\, II$ solutions, built from the structure tensor density \eqn{structCstSII}.
\par\noindent To diagonalise and put into the standard form
\begin{equation}\label{snBWIII}
n_{(s)VIII}= diag.(-1,1,1)
 \end{equation}
the structure tensor density \eqn{nIII} we just have to determine the eigenvalues and the appropriately rescaled eigenvectors of the matrix $n_{III}^{\alpha\,\beta}$ . The eigenvalues are given by the three real, distinct (and non vanishing) roots $\lambda^{(p)}$ of the polynomial
\begin{equation}
\lambda^3+a\,\lambda^2-(a^2+2)\,\lambda-a^3=0\qquad.
\end{equation}
For positive values of $a$, we always have two eigenvalues negative and one positive ; it is the converse for negative $a$. 
\par\noindent The corresponding eigenvectors can be  chosen proportional to
\begin{equation}
u^{(p)}_\alpha=((a+\lambda^{(p)})^2-1,\,(a+\lambda^{(p)}),\, 1)\qquad .
\end{equation}
They may be used to diagonalise $n_{III}^{\alpha\,\beta}$ and lead, at an intermediate step, to the diagonal matrix :
\begin{eqnarray}
{\tilde n}^{p\,q}&=&-\lambda^{(p)}\frac {u^{(p)}\cdot u^{(q)}}{(\lambda^{(1)}-\lambda^{(2)})(\lambda^{(2)}-\lambda^{(3)})(\lambda^{(3)}-\lambda^{(1)})}\nonumber\\
&=& \delta^{p\,q}\,\lambda^{(p)}\frac{(4\,a^2+1)(\lambda^{(p)})^2+4\,a(2\,a^2+1)\,\lambda^{(p)}+(4\,a^4-a^2+2)}{2\sqrt{16\,a^4+13\,a^2+8}}
=:\tilde n^{(p)}\,\delta^{p\,q}\quad .
\end{eqnarray}
To obtain the standard expression of the structure constant density tensor we still have to rescale this matrix. Assuming that we ordered the roots such that $\lambda^{(1)}>\lambda^{(2)}>\lambda^{(3)}$ when $a>0$ or $\lambda^{(1)}<\lambda^{(2)}<\lambda^{(3)}$ when $a<0$, this is achieved thanks to the transformation defined by the diagonal matrix $L$ of components~:
\begin{equation}
L^{p\,q}:=-\mbox{sgn} (a)\, \delta^{p\,q}\,\left( \frac {\vert a\vert^3}{2\vert {\tilde n}^{(p)}\vert\sqrt{16\,a^4+13\,a^2+8}}\right)^{1/2}=:\delta^{p\,q}\,L^{(p)}\qquad .
\end{equation}
Finally the metric components
 with respect to the usual right invariant  one--forms on the Bianchi $VIII$ group :
\begin{equation}\label{B8of}
\theta^1=\mathrm{d}t-\sinh x\,\mathrm{d}y\quad,\quad \theta^2=\cos t\,\mathrm{d}x-\sin t\cosh x\,\mathrm{d}y\quad,\quad\theta^3=\sin t\,\mathrm{d}x+\cos t\cosh x\,\mathrm{d}y\quad ,
\end{equation}
are given by the matrix product 
\begin{equation}
g_{p\,q}=\Lambda_p^\alpha\,\Lambda_q^\beta\,\eta_{\alpha\,\beta}\qquad \mbox{with}\qquad (\Lambda^{-1})^p_\alpha=
L^{(p)}\,u^{(p)}_\alpha\qquad .
\end{equation}
Let us emphasise that this metric will not be diagonal and cannot be diagonalised\footnote{In other words the solution \eqn{III.GMG} cannot be 
obtained from a diagonal ansatz  like
$$
\mathrm{d}s^2=A\,(\theta^1)^2+B\,(\theta^2)^2+C\,(\theta^3)^2\qquad, 
$$
which always leads to metrics of Petrov type $D$.} by an $Iso(1,2)$ transformation {  (being itself also of ``type $II$")}, but put into the form :
\begin{equation}
\mathrm{d}s^2=a^{-2}(-(\theta^1)^2+ (\theta^2)^2+(\theta^3)^2) +\gamma\,(\theta^1+ \theta^3)\,\theta^2\qquad,
\end{equation}
where $\gamma\neq 0$ is an arbitrary parameter.
{  \par\noindent Similarly solutions (\eqn{II.2}, \eqn{II.3} and \eqn{II.4}) can be written, in terms of the one-forms 
\eqn{B8of}, as :
 \begin{equation}
\mathrm{d}s^2=g^{II}_{\alpha\,\beta}\,\theta^\alpha\,\theta^\beta\qquad,
\end{equation}
where the matrix of component $(g^{II}_{\alpha\,\beta})$ is numerically equal to the matrix $(n_{II}^{\alpha\,\beta})$ that was introduced in Eq.\eqn{tensor.II}; we also demand the constraint $b\,a^2=1$ upon its elements.\\
Furthermore, the solution \eqn{L.4} can be written,
 \begin{equation}
\mathrm{d}s^2=|c|\,(\theta^1)^2-\frac{2\,\theta^2\,\theta^3}{\sqrt{|c|}}\qquad,
\end{equation}
in terms of the one-forms defining Bianchi spaces $VI_h$ or $VII_h$ respectively \cite{SKMHH} }:
 \begin{eqnarray}
&&\theta^1=e^{-h\,z}\,\left(\cosh z\,\mathrm{d}x+\sinh z\,\mathrm{d}y\right)\,,\quad
\theta^2=e^{-h\,z}\,\left(\sinh z\,\mathrm{d}x+\cosh z\,\mathrm{d}y\right)\,,\quad
\theta^3=\mathrm{d}z\qquad , \\
&&\theta^1=e^{-h\,z}\,\left(\sin z\,\mathrm{d}x+\cos z\,\mathrm{d}y\right)\,,\quad
\theta^2=e^{-h\,z}\,\left(\cos z\,\mathrm{d}x-\sin z\,\mathrm{d}y\right)\,,\quad
\theta^3=\mathrm{d}z\qquad . 
\end{eqnarray}

Let us now illustrate the analytical approach.
The $S\,II$ structure constants correspond to a group admitting an abelian two-dimensional subgroup. 
The corresponding  Cartan equations  are obtained from the structure tensor density \eqn{structCstSII} 
and the space like structure vector $k_\alpha=(0,0,k)$. They read :
\begin{equation}
[\xi_1,\,\xi_2]=0\qquad,\qquad[\xi_1,\,\xi_3]=(k+\nu)\xi_1+(\nu-a)\xi_2\qquad,\qquad[\xi_2,\,\xi_3]=-(a+\nu)\xi_1+(k-\nu)\xi_2\quad.
\end{equation}
Thanks to the presence of the abelian subgroup, the integration of these equations is immediate (see ref. \cite{GbS} 
for a discussion of this problem in the framework of the Bianchi $III$ group) and,  after a specific choice of the integration constants, leads to :
\begin{equation}
\xi_1=\partial_x\quad,\quad\xi_2=\partial_y\quad,\quad\xi_3=\partial_z+r(x,y)\partial_x+s(x,y)\partial_y\quad,
\end{equation}
where the functions $r(x,y)$ and $s(x,y)$ are linear :
\begin{equation}
r(x,y)=(k+\nu)\,x-(a+\nu)\, y\qquad\text{and}\qquad s(x,y) =( \nu-a)\,x+(k-\nu) \, y\qquad.
\end{equation}
We immediately obtain the dual right invariant one-forms that define the metric :
\begin{equation}
\theta^1=\mathrm{d}x-r(x,y)\,\mathrm{d}z\quad,\quad\theta^2=\mathrm{d}y-s(x,y)\,\mathrm{d}z\quad,\quad\theta^3=\mathrm{d}z\quad .
\end{equation}
The metric reads
\begin{equation}\label{ppwAdSmet1}
\mathrm{d}s^2=-\left(\mathrm{d}x-r(x,y)\,\mathrm{d}z\right)^2+\left(\mathrm{d}y-s(x,y)\,\mathrm{d}z\right)^2+\mathrm{d}z^2
\end{equation}
and solves the massive gravity equations when the algebraic conditions \eqn{SIITMG}, \eqn{SIINMG}, \eqn{SIIGMG} are satisfied. 

\par\noindent The metric \eqn{ppwAdSmet1} has an explicit one--parameter (hereafter labelled as $\lambda_3$) isometry subgroup :
\begin{equation}\label{lam3}
z\mapsto z+\lambda_3
\end{equation}
 but  hides the two--parameter abelian isometry subgroup given by
\begin{equation}
x\mapsto x+  p(z)\qquad,\qquad y\mapsto y+  q(z)\qquad,\qquad z\mapsto z
\end{equation}
where $p(z)$ and $q(z)$ are the solutions (depending on two arbitrary constants denoted hereafter $\lambda_1$ and $\lambda_2$ : 
the group parameters) of the differential system :
\begin{eqnarray}
&&p'(z)=(k+\nu)\,p(z)-(a+\nu)\,q(z)\qquad,\\
&&q'(z)=(\nu-a)\,p(z)+(k-\nu)\,q(z)\qquad ,
\end{eqnarray}
whose solution reads
\begin{eqnarray}
&&p(z)= \lambda_1\,e^{k\,z}\left( a\,\cosh(a\,z)+\nu\,\sinh(a \,z)\right)-\lambda_2\,e^{k\,z}\,(\nu+a)\,\sinh(a\,z)\\
&&\phantom{f(z)}:=\lambda_1 p_1(z)+\lambda_2\,p_2(z)\qquad,\label{lam12f}\\
&&q(z)= \lambda_1\,e^{k\,z}\,(\nu-a )\,\sinh(a\,z)+\lambda_2\,e^{k\,z}\,\left( a\,\cosh(a\,z)-\nu\,\sinh(a \,z)\right)\\
&&\phantom{q(z)}:=\lambda_1\,q_1(z)+\lambda_2\,q_2(z)\label{lam12g}\qquad.
\end{eqnarray}
Now it is immediate to write the most general left invariant vector, a Killing vector of the metric \eqn{ppwAdSmet1} 
depending on the three parameters $\lambda_1,\lambda_2$ and $\lambda_3$ introduced respectively in eqs \eqn{lam12f} or \eqn{lam12g} and \eqn{lam3}   :
\begin{equation}
\tilde{\xi} = \lambda_3\,\partial_z+p(z)\,\partial_x+q(z)\,\partial_y\qquad .
\end{equation}
To make explicit the action of the abelian two-dimensional subgroup we have to use the group parameters as coordinates. 
Assuming $a\neq 0$, we introduce new coordinates $X$ and $Y$  which are defined by
\begin{eqnarray}
&&x=p_1(z)\,X+p_2(z)\,Y\qquad,\\
&&y=q_1(z)\,X+q_2(z)\,Y\qquad,
\end{eqnarray}
and we obtain
\begin{eqnarray}
&&\theta^1=p_1(z)\,\mathrm{d}X+p_2(z)\,\mathrm{d}Y\,\nonumber \\
&&\phantom{\theta^1}=\frac{e^{k\,z}}{2}\left\{e^{a\,z}\left(\nu+a\right)\left(\mathrm{d}X+\mathrm{d}Y\right)+e^{-a\,z}
\left(\left(a-\nu\right)\,\mathrm{d}X+\left(a+\nu\right)\,\mathrm{d}Y\right)\right\}\,\qquad,\\
&&\theta^2=q_1(z)\,\mathrm{d}X+q_2(z)\,\mathrm{d}Y\,\nonumber \\
&&\phantom{\theta^2}=\frac{e^{k\,z}}{2}\left\{e^{a\,z}\left(\nu-a\right)\left(\mathrm{d}X-\mathrm{d}Y\right)+e^{-a\,z}
\left(\left(a-\nu\right)\,\mathrm{d}X+\left(a+\nu\right)\,\mathrm{d}Y\right)\right\}\qquad.
\end{eqnarray}
A last coordinate transformation:
\begin{equation}
X\mapsto\frac{\sigma\,\nu\,v+k^2\,(\nu+a)\,u}{2\,k^2\,a\,\sqrt{\vert a \vert}}\qquad,\qquad Y\mapsto  \frac{\sigma\,\nu\,v+k^2\,(\nu-a)\,u}{2\,k^2\,a\,\sqrt{\vert a \vert}}\qquad,\qquad 
z\mapsto -\frac 1 k\ln\,\zeta\qquad,
\end{equation}
where $\sigma:=-\text{sgn}(a)\,\nu$, provides the usual expression of the 
so-called pp-wave AdS 
\cite{GPS,  Moutsopoulos:2012bi}] metric or null warped AdS metric : 
 \begin{equation}\label{ppwAdSmet2}
ds^2=\frac 1 {k^2}\,\frac{\mathrm{d}u\,\mathrm{d}v+\mathrm{d}{\zeta}^2}{\zeta^2}+\sigma\,\frac{\mathrm{d}u^2}{\zeta^{2(a/k +1)}}\qquad.
\end{equation}
Obviously the coordinate system $(u,v,\zeta)$ (or equivalently $(x,y,z)$) defines a local chart but does not cover the whole manifold.

\subsection{Non simply transitive groups}

Apart from the aforementioned cases in subsection  \ref{Sol.trans}, there  is also a special class  homogeneous spaces: the Kantowsky--Sachs  spacetimes \cite{Sachs}  that do not admit a simply-transitive three dimensional isometry group. The corresponding one-parameter metric describes homogeneous spaces of the form $R\times S^2$, on which acts (multi-transitively) a 4-parameter isometry group that does not contain any 3-parameter transitive subgroup:
\begin{equation}
\mathrm{d}s^2=-\mathrm{d}t^2+R^2\,(\mathrm{d}\theta^2+\sin^2\theta\,\mathrm{d}\phi^2)\qquad .
\end{equation}
{This geometry is conformally flat, with traceless Ricci and $\widehat  K_{\alpha\,\beta}$ tensors of Petrov type
$D_t$.}
Thus, it can utmost satisfy the NMG (and so the GMG) field equations. It turns out that the latter are satisfied for arbitrary $\mu$ but $\Lambda=\xi=\frac{1}{2R^2}\,$.
Let us emphasise that the hyperbolic or the flat version of this metric (with $\sin\theta$ is replaced by 
$\sinh\theta$ or $\theta$ respectively) 
admit   3-parameter  transitive isometry groups. Indeed the   $SO(3)$ group only admits one dimensional subgroups, whereas  in the framework of the hyperbolic version of the metric, the Lorentz group $SO(1,2)$ possesses two dimensional subgroups. Thus, the latter geometry has  already been considered in the previous subsections.
{The hyperbolic metric  is conformally flat, with traceless Ricci and $\widehat K_{\alpha\,\beta}$ tensors of Petrov type $D_t$. It solves the equations NMG (and GMG) with arbitrary $\mu$ but $\Lambda=\xi=-\frac{1}{2R^2}$. Actually it appears as a special solution of Bianchi type $III$, obtained from a $S\, III$ non-unimodular Lie algebra (\ref{structCstSIII}) with $a=0$ and $k=-\nu$,   given by Eq.(\ref{KSHNMG}) and Eqs (\ref{KSH1GMG}, \ref{KSH2GMG}) for NMG and GMG respectively.}

\section{Vanishing scalar invariant geometries}\label{VSIsec}
\label{VSI}
{ In this section we complete the list of Lorentzian  spacetimes with constant scalar invariant geometries  solving the MG field equations  by examining  the solutions provided by vanishing scalar invariant (VSI) geometries.}

All the homogeneous geometries considered here above share the common property that all their scalar geometrical objects 
 are constants. 
Notably, in the framework  of Lorentzian geometries, there exist spaces (VSI spaces) which are not locally homogeneous 
but  have  all scalar invariants built out of 
their curvature tensors vanishing. These geometries constitute a subclass of the Kundt geometries considered in  \cite{CPS2} and are explicitly known in three dimensions \cite{CHP}. Their
metrics are characterised by the existence of a null geodesic vector field (which in three dimensions  implies vanishing of shear and twist).
If we exclude flat space, there are two possible such metrics\footnote{In ref. \cite{CHP} four non flat expressions of the metric are displayed, 
but the last two (denoted as D1\&F1) are special cases of the first ones.} (labelled {\bf A1} and {\bf B1} in ref. \cite{CHP}) that can be written as follows
\begin {equation}\label{metVSI}
ds^2=-2\,\mathrm{d}u\,\left[\mathrm{d}v+ \frac 12 F(u,v,x) \,\mathrm{d}u+ W(u,v,x)  \, \mathrm{d}x\right]+ \mathrm{d}x^2
\end{equation}
with special expressions of the function $F(u,v,x)$ and $W(u,v,x)$.
Using a null frame $\{l ,n,m\}$ such that $l^\alpha n_\alpha=-1=-m^\alpha m_\alpha$, these metrics lead to    Ricci tensors that read as
\begin {equation}
R_{\alpha\,\beta}=\phi\,l_\alpha l_\beta\qquad\text{or}\qquad R_{\alpha\,\beta}=\psi\,l_{(\alpha}m_{\beta)}
\end{equation}
with $\phi\neq0$ and $\psi\neq0$ .   The first one is of Petrov type $N$, whereas the second one is of Petrov type $III$.
Of course the functions appearing in the  metric \eqn{metVSI} can be modified by coordinate transformations that preserve  their writing :
\begin{equation}
\label{coo.tran}
u\mapsto {\cal U}[\tilde u]\qquad,\qquad v\mapsto \frac{\tilde v}{\dot{\cal U}[\tilde u]}+{\cal F}[\tilde u,\,\tilde x]\qquad,\qquad x\mapsto \tilde x+{\cal G}[\tilde u]\qquad .
\end{equation}
Using these transformations it is easy to check that the induced transformations on the functions $(F,W)$ read
\begin{equation}
\label{ind.tran}
\widetilde F=F\, \dot {\cal U}^2+2\left(\dot{\cal F}\,\dot {\cal U}+W\,\dot {\cal G}\,\dot {\cal U}-\tilde v\frac{\ddot {\cal U}}{\dot {\cal U}}\right)-\dot {\cal G}^2\qquad,\qquad
\widetilde W=W+{\cal F}'-\dot {\cal G}\qquad,
\end{equation}
where $(F,W)$ are expressed in terms of the  $(\tilde u,\tilde v,\tilde x)$ coordinates via their $(u, v, x)$ dependence given by  
\eqn{coo.tran}. Dot and prime denote partial derivatives with respect to the new coordinates $\tilde u$ and $\tilde x$.
In what follows we   describe the resolution of the massive gravity field equations on VSI spaces. Let us notice that for consistency 
we have to assume a zero cosmological constant, as all scalar invariants of  VSI metrics vanish. \\
We have solved the equations of TMG, NMG and GMG for type {\bf A1} and {\bf B1}. It turns out that the only nontrivial equations
are those corresponding the $[u,x]$ and $[u,u]$ components. We start by considering the simplest one, namely the $[u,x]$ equation,  plug its solution into the  $[u,u]$ one and solve it.

\begin{itemize}
\item{Type {\bf A1}}\\
Here the metric components are a priori of the form
\begin{equation}
\label{A1}
F(u,v,x)=v\,f_1(u,x)+f_0(u,x)\qquad,\qquad W(u,v,x)=w_0(u,x)\qquad.
\end{equation}
Before proceeding to the equations of motion we shall first discuss the gauge fixings that are allowed by the coordinate transformations
\eqn{coo.tran}. At first we note that we can always eliminate $w_0(u,x)$ with an
appropriate choice of ${\cal F}$
\begin{equation}
{\cal F}(\tilde u,\tilde x)=-\int\,w_0(\tilde u,\tilde x)\,{\mathrm d}\tilde x+\dot{\cal G}(\tilde u)\tilde x+{\cal H}(\tilde u)\quad;
\end{equation}
thus we consistently assume that $w_0(u,x)=0$.
Using the latter, it easy to check that the residual coordinate transformations induce the transformations 
\begin{eqnarray}
\label{ind.tran.A1}
&&\tilde f_1=f_1\,\dot{\cal U}-2\frac{\ddot{\cal U}}{\dot{\cal U}}\qquad,\\
&&\tilde f_0=f_0\,\dot{\cal U}^2+\left(\tilde x\,\dot{\cal G}+{\cal H}\right)f_1\,\dot{\cal U}^2+
2\left(\tilde x\,\ddot{\cal G}+\dot{\cal H}\right)\,\dot{\cal U}-\dot{\cal G}^2\qquad.\nonumber
\end{eqnarray}
Hereafter we shall present the solution of the field equations in a fixed coordinate system, where we have eliminated as many 
as possible gauge functions in the expression of the metric.

\subitem{$\star$ TMG}\\
The only nontrivial field equations are $[u,x]$ and $[u,u]$. The first one gives
\begin{equation}
\partial^2_x f_1(u,x)+\mu\,\partial_xf_1(u,x)=0\qquad,
\end{equation}
whose general solution reads
\begin{equation}
\label{A1-f1-TMG}
f_1(u,x)= c_0(u)+c_1(u)\,e^{-\mu\,x}\qquad.
\end{equation}
Using \eqn{ind.tran.A1} we can eliminate $c_0(u)$ and set (locally) $c_1(u)$ to one
by choosing $2\,\ddot {\cal U}/\dot {\cal U}^2=c_0(u)$\,, $c_1(u)\,e^{-\mu{\cal G}}\,\dot{\cal U}=1$, {\it i.e.} in 
Eq. \eqn{ind.tran.A1} the $u={\cal U}(\tilde u)$ coordinate transformation as the inverse transformation of
\begin{equation}
\tilde u=\frac{1}{c_\star}\int e^{-\frac{1}{2}\,\int c_0(u)\,\mathrm{d}u}\,\mathrm{d}u\qquad ,\qquad c_\star=\const
\end{equation}
 and with an appropriate sign for $c_\star$
 \begin{equation}
 {\cal G}(\tilde u)=\frac{1}{\mu}\left(\frac{1}{2}\int c_0(u)\,\mathrm{d}u+\ln[c_\star\,c_1({\cal U}(\tilde u))]\right)\qquad .
 \end{equation}
  Let us notice that if $c_1(u)=0$ the metric is flat.
Using the latter solution, the $[u,\, u]$ equation reduces to
\begin{equation}
 \partial^3_xf_0(u,x)+\mu\, \partial^2_xf_0(u,x)=-\frac{\mu}{2}\,e^{-2\mu x}\qquad,
\end{equation}
whose general solution reads
\begin{equation}
\label{A1-f0-TMG}
 f_0(u,x)=q_0(u)+q_1(u)\,x+q_2(u)\,e^{-\mu\,x}+\frac{e^{-2\mu\,x}\,}{8\,\mu^2}\qquad.
\end{equation}
Last but not least, we can also eliminate $q_2(u)$ with
the appropriate choice of ${\cal H}(\tilde u)=-q_2(\tilde u)$.

\subitem{$\star$ NMG} \\
Working similarly as in the TMG we obtain
\begin{equation}
\label{A1-f1-NMG}
f_1(u,x)= e^{-\sqrt{\xi}\,x}+c_2(u)\,e^{\sqrt{\xi}\,x}
\end{equation}
and
\begin{eqnarray}
\label{A1-f0-NMG}
f_0(u,x)&=&
q_0(u)+q_1(u)x+q_2(u)\,e^{-x\,\sqrt{\xi}}+\frac{2x\sqrt{\xi}-5}{2\xi}\,c_2'(u)\,e^{x\,\sqrt{\xi}}\nonumber\\
&&+\frac{1}{6\xi}\,e^{-2x\,\sqrt{\xi}}+\frac{c_2^2(u)}{6\xi}\,e^{2x\,\sqrt{\xi}}\qquad.
\end{eqnarray}

\subitem{$\star$ GMG}\\ 
In the same way we find
\begin{eqnarray}
\label{A1-f1-GMG}
f_1(u,x)=\left\{
	\begin{array}{ll}
	e^{\lambda_-\,x}+c_2(u)\,e^{\lambda_+\,x}~,  & \mbox{where}\ \lambda_\pm=\frac{\xi\pm\sqrt{\xi(\xi+4\mu^2)}}{2\mu}~, \\
	c_2(u)\,x\,e^{-2\mu\,x}~, & \mbox{if } \xi=-4\mu^2
	\end{array}
\right.
\end{eqnarray}
and 
\begin{eqnarray}
\label{A1-f0a-GMG}
&&f_0(u,x)=q_0(u)+r_0(u,x)+(q_1(u)+r_1(u,x))\,x+r_2(u,x)e^{\lambda_-x}+(q_3(u)+r_3(u,x))e^{\lambda_+x}~,\nonumber\\
&&r'_0(u,x)=\frac{1+\mu\,x}{\xi\,\mu}\,j(u,x)\,,\quad r_1'(u,x)=-\frac{j(u,x)}{\xi}\,,\quad 
r_{2|3}'(u,x)=\mp\frac{e^{-x\lambda_\mp}\,j(u,x)}{\lambda^2_\mp(\lambda_+-\lambda_-)}~,
\end{eqnarray}
whereas for $\xi=-4\mu^2$
\begin{eqnarray}
\label{A1-f0b-GMG}
&&f_0(u,x)=q_0(u)+r_0(u,x)+(q_1(u)+r_1(u,x))\,x+\left\{r_2(u,x)+(q_3(u)+r_3(u,x))x\right\}e^{-2\mu\,x}~,\nonumber\\
&&r_0'(u,x)=-\frac{1+\mu\,x}{4\mu^3}\,j(u,x)\,, \qquad r_1'(u,x)=\frac{j(u,x)}{4\mu^2}\,, \qquad 
r_2'(u,x)=\frac{(1-\mu\,x)\,e^{2x\,\mu}\,j(u,x)}{4\mu^3}\,,\nonumber\\
&&r'_3(u,x)=\frac{e^{2\mu\,x}\,j(u,x)}{4\mu^2}~,
\end{eqnarray}
where 
\begin{equation}
j(u,x):=\left(\left(\partial_xf_1(u,x)\right)\partial_x+ f_1(u,x)\left(\partial_x^2-\frac{\xi}{2\mu}\partial_x\right)-\frac{\xi}{\mu}\partial_x\partial_u+2\partial_u\partial_x^2\right)f_1(u,x)\qquad.
\end{equation}

\item{Type {\bf B1}}\\
The metric components are given by 
\begin{equation}
\label{B1}
F(u,v,x)=-\frac{v^2}{x^2}+ v\,f_1(u,x)+f_0(u,x)\qquad,\qquad W(v,u,x)=-\frac{2v }{x }+w_0(u,x)\qquad.
\end{equation}
Applying the coordinate transformations \eqn{coo.tran}, in this case we find that the 
ones preserving the form of $W$ require $u=\tilde u+u_0$, $x=\tilde x+x_0$ (where $u_0$ and $x_0$ are constants), while ${\cal F}$ remains an arbitrary function. Fixing it as follows : 
\begin{equation}
\left(\partial_x-\frac{2}{x}\right){\cal F}(u,x)=
-w_0(u,x)\,\Longrightarrow {\cal F}(u,x)=-x^2\left(\int_x\frac {w_0(u,x')}{x'^2}\,\mathrm{d}x'\, -{\cal H}(u)\right)\qquad.
\end{equation}
allows to eliminate $w_0(u,x)$. Thus we can set consistently $w_0(u,x)=0$ 
and obtain 
\begin{eqnarray}
\label{ind.tran.B1}
&&\widetilde F=-\frac{\tilde v^2}{\tilde x^2}+\tilde v\,\tilde f_1+\tilde f_0\qquad,\\
&&\tilde f_1=f_1-2{\cal H}\qquad,\qquad \tilde f_0=f_0+\tilde x^2{\cal H}(f_1-1)+2\tilde x^2\dot{\cal H}\qquad.\nonumber
\end{eqnarray}
As for Type {\bf A1} we shall present the solutions of the field equations in a coordinate system fixed by eliminating as 
many as possible gauge functions in the expression of the metric. 
\subitem{$\star$ TMG}\\
The only nontrivial field equations are $[u,x]$ and $[u,u]$. The first one gives
\begin{equation}
\left(\partial^2_x+\left(\mu+\frac{1}{x}\right)\partial_x\right)f_1(u,x)=0\qquad,
\end{equation}
whose solution reads
\begin{equation}
\label{B1-f1-TMG}
f_1(u,x)=c_0(u)+c(u)\,Ei(-\mu x)\qquad,
\end{equation}
where $Ei(z)= -{\cal PV}\int_{-z}^{ +\infty} e^{-t}/t\,\mathrm{d}t\,,$ denotes the exponential integral function. 
Using \eqn{ind.tran.B1} we can eliminate, for example, $c_0(u)$ by choosing $c_0(u)=2\,{\cal H}(u)$.
Using the latter equation, the $[u,\,u]$ equation reads
\begin{eqnarray}
&&\left(\partial^3_x+\left(\mu-\frac{3}{x}\right)\left(\partial^2_x-\frac{2}{x}\partial_x+\frac{2}{x^2}\right)\right)f_0(u,x)=j(u,x)\qquad,\nonumber\\
&&j(u,x):=\frac{e^{-\mu\,x}}{2x}\left(c^2(u)Ei(-\mu\,x)+2c'(u)\right)\qquad,
\end{eqnarray}
whose solution is given by :
\begin{eqnarray}
\label{B1-f0-TMG}
&&f_0(u,x)=x\left(q_0(u)+r_0(u,x)+(q_1(u)+r_1(u,x))\,x+(q_2(u)+r_2(u,x))\,e^{-\mu\,x}\right)~,\\
&&r_0'(u,x)=-\frac{(1+\mu\,x)j(u,x)}{\mu^2\,x}\, , \quad r_1'(u,x)=\frac{j(u,x)}{\mu\,x}\, , \quad r_2'(u,x)=\frac{e^{\mu\,x}j(u,x)}{\mu^2\,x}~.\nonumber
\end{eqnarray}

\subitem{$\star$ NMG}\\
Working similarly as in the TMG we obtain
\begin{equation}
\label{B1-f1-NMG}
f_1(u,x)=c_1(u)\,Ei(-x\sqrt{\xi}\,)+c_2(u)\,Ei(x\sqrt{\xi}\,)
\end{equation}
and
\begin{eqnarray}
\label{B1-f0-NMG}
&&f_0(u,x)=x\left(q_0(u)+r_0(u,x)+(q_1(u)+r_1(u,x))\,x+(q_2(u)+r_2(u,x))\,e^{-x\sqrt{\xi}}+(q_3(u)+r_3(u,x))\,e^{x\sqrt{\xi}}\right)\qquad,\nonumber\\
&&r_0'(u,x)=\frac{j(u,x)}{x}\qquad,\qquad r_1'(u,x)=-\frac{j(u,x)}{x\xi}\qquad,\qquad 
r_{2|3}'(u,x)=\mp\frac{e^{\pm x\,\sqrt{\xi}}\,j(u,x)}{2x\,\xi^{3/2}}\qquad,
\end{eqnarray}
where 
\begin{equation}
 j(u,x):=\left(\left(\partial_xf_1(u,x)\right)\partial_x +2\partial_u\partial_x^2+f_1(u,x)\partial^2_x\right)f_1(u,x)\qquad.
\end{equation}

\subitem{$\star$ GMG}\\
As in the TMG, we obtain 
\begin{eqnarray}
\label{B1-f1-GMG}
f_1(u,x)= \left\{
	\begin{array}{ll}
		c_1(u)\,Ei(\lambda_-x)+c_2(u)\,Ei(\lambda_+x)~,  & \mbox{where}\ \lambda_\pm=\frac{\xi\pm\sqrt{\xi(\xi+4\mu^2)}}{2\mu}~, \\
		c_1(u)\,e^{-2\mu\,x}+c_2(u)\,Ei(-2\mu\,x)~, & \mbox{if } \xi=-4\mu^2
	\end{array}
\right.
\end{eqnarray}
and 
\begin{eqnarray}
\label{B1-f0a-GMG}
&&f_0(u,x)=x\left(q_0(u)+r_0(u,x)+(q_1(u)+r_1(u,x))\,x+(q_2(u)+r_2(u,x))\,e^{\lambda_-\,x}+(q_3(u)+r_3(u,x))\,e^{\lambda_+\,x}\right)~,\nonumber\\
&&r'_0(u,x)=\frac{1+\mu\,x}{\xi\,\mu\,x}\,j(u,x)\qquad,\qquad r_1'(u,x)=-\frac{j(u,x)}{\xi\,x}\qquad,\qquad 
r_{2|3}'(u,x)=\mp\frac{e^{-x\lambda_\mp}\,j(u,x)}{\lambda^2_\mp(\lambda_+-\lambda_-)\,x}~,
\end{eqnarray}
whereas for $\xi=-4\mu^2$
\begin{eqnarray}
\label{B1-f0b-GMG}
&&f_0(u,x)=x\left(q_0(u)+r_0(u,x)+(q_1(u)+r_1(u,x))\,x+\left\{q_2(u)+r_2(u,x)+(q_3(u)+r_3(u,x))x\right\}e^{-2\mu\,x}\right)~,\nonumber\\
&&r_0'(u,x)=-\frac{1+\mu\,x}{4\mu^3\,x}\,j(u,x)\,, \qquad r_1'(u,x)=\frac{j(u,x)}{4\mu^2\,x}\,, \qquad 
r_2'(u,x)=\frac{(1-\mu\,x)\,e^{2x\,\mu}\,j(u,x)}{4\mu^3\,x}\,,\nonumber\\
&&r'_3(u,x)=\frac{e^{2\mu\,x}\,j(u,x)}{4\mu^2\,x}~,
\end{eqnarray}
where 
\begin{equation}
j(u,x):=\left(\left(\partial_xf_1(u,x)\right)\partial_x +f_1(u,x)\left(\partial_x^2-\frac{\xi}{2\mu}\partial_x\right)-
\frac{\xi}{\mu}\partial_x\partial_u+2\partial_u\partial_x^2\right)f_1(u,x)\qquad.
\end{equation}

\end{itemize}

\section{Discussion and conclusion}

To summarise, we have obtained all locally homogeneous spaces, solutions of the TMG, NMG and GMG field equations (with cosmological constant), at least formally. 
We have classified them according to  canonical representations of the structure constants obtained by Lorentz transformations, and for anti-de Sitter geometries we have discussed the appearance of a tachyonic massive mode of the graviton. To obtain the explicit expressions 
of the metrics, we still have to solve an elementary algebraic problem that consists of finding the linear transformation that maps the expressions of the 
structure tensor density and the structure vector from which we start on their usual canonical expressions, i.e. express the one--forms defining our coframe  as linear combination of the canonical ones.  We may also directly integrate the expressions of the right invariant forms. Both approaches are 
explicitly illustrated in subsection \ref{coord.repr.SII}, {  for solutions with structure tensor densities of} types $III$ and $SII$.

We have also determined the Petrov types of the traceless Ricci tensor, the Cotton--York tensor and the traceless $K_{\mu\nu}$ tensor
of all Lorentzian three-dimensional homogeneous geometries, which proves to be a useful tool to recognize equivalent solutions: 
for instance solutions of TMG, which are of Petrov type D, 
are biaxially squashed $AdS_3$ geometries \cite{CPS1}.

In brief, we found solutions with structure tensor densities of type $I$ that correspond to all unimodular Bianchi types and allow all values of the cosmological constant; 
for those structure tensor densities of type $II$ (which could lead  to the Bianchi types $II$, $VI_0$, $VII_0$ and $VIII$) there are solutions of Petrov types $N$ and  $II$; in case structure tensor densities of type $III$ ( corresponding to Bianchi type $VI_0$ and $VIII$) only the GMG has  solutions of Petrov type $III$ with negative cosmological constant, which was not previously known;
for structure tensor densities of type $IV$ ( corresponding to Bianchi type $VI_0$ and $VIII$) there are solutions of  Petrov type $I_{\mathbb{C}}$ and a solution of Petrov type $D_s$ in the case of 
GMG.\\
 In the case of non-unimodular homogeneous spaces, the Lorentzian
classification of the algebras leads to solutions for  structure tensor densities of types $T$ and $SI$  (corresponding to all non-unimodular types Bianchi
 types $III$, $IV$, $VI_h$ and $VII_h$) that allow both signs of the cosmological constant;
for structure tensor densities of type $S\,II$ (corresponding to Bianchi types  $IV$, $III$ and $VI_h$) there are solutions with only negative values of the cosmological constant;
for structure tensor densities of type $S\,III$ (corresponding to Bianchi types $III$ and $VI_h$) both signs of the cosmological constant are allowed;
for structure tensor densities of type $L$ ( corresponding to all non-unimodular Bianchi types ) there are new solutions of Petrov type $II$ with 
positive cosmological constant
for TMG and NMG and negative one for the GMG; for structure tensor densities of type $L'$ (corresponding to Bianchi types $III,V,VI_h$) the only possible solution 
is flat space.  
 
In addition, we have also  obtained the solutions of TMG, NMG and GMG field equation for VSI geometries (which imply a vanishing cosmological constant).
The solutions  that  we found are of Petrov type $III$. After having fixed the coordinate system, they all contain several arbitrary functions of a lightlike coordinate, 
reflecting the physical degrees of freedom of these solutions. 

{ Regarding} the solutions we obtained, let us make two more remarks. Firstly, we recovered, as structure tensor densities types $I$, $SI$ and $SIII$  (and Petrov type $D_t$) 
the one-parameter family of deformed anti-de Sitter geometries \cite{RoSp}  that  includes the famous G\"odel geometry \cite{God}. On the other hand, 
each of the VSI solutions,  which were given in section \ref{VSIsec}, contain various arbitrary functions. As it is known, these functions reflect 
the wave propagating aspect of these solutions, and describe the arbitrariness of the profile of these waves.

Of course there are numerous known solutions among the ones  that  we obtained. Nevertheless (to our knowledge) some of them are new. In particular,
we found CSI solutions of Petrov type $II$ and $III$ in Eqs \eqn{II.2},\eqn{II.3},\eqn{II.4},\eqn{III.GMG},\eqn{L.4} and the VSI  spacetimes in section \ref{VSI}. 
However, the  main goal of our analysis was to make a complete list of all the locally homogeneous and VSI geometries  that  constitute CSI  spacetimes. 
From a physical perspective such solutions are among the simplest to study, and may also contain fruitful results  that  deserves further study.

The above results provide a classification of all the solutions of massive gravity theories on CSI geometries. Of course not all of them can be considered as 
classical background geometries. For instance, it is well known that on Bianchi IX solutions, due to the compactness of the space, no global causal structure 
could be defined. Some of the squashed anti-de Sitter metrics  that  were obtained may also suffer from causality pathologies \cite{RoSp}. Moreover we insist on the 
fact that the solutions  that  were obtained are often only local solutions, providing geodesically incomplete spaces. Thus, it would be very interesting and instructive to study the perturbative stability and the global structure of all these backgrounds, and to identify all the 
physically relevant configurations. It also would be newsworthy to check if the Petrov type $D$ solutions of NMG and GMG
are biaxially squashed $AdS_3$, like as in the TMG \cite{CPS1}.

\section*{Acknowledgments}
We would like to thank E. Bergshoeff, { P. Cartier}, T. Damour, S. Detournay, { S. Katmadas}, U. Moschella and { P. M. Petropoulos}  for interesting and useful comments and discussions on various of aspects of this work.
This work has been supported by  ``Actions de recherche concert\'ees (ARC)'' de la Direction g\'en\'erale de l'Enseignement non obligatoire et de la Recherche scientifique Direction de la Recherche scientifique Communaut\'e fran\c{c}aise de Belgique, and by IISN-Belgium (convention 
4.4511.06).
Ph.\,S. and K.\,S. reiterate their thanks to IH\'ES and  the University of Patras respectively, 
for hospitality where part of this work was developed. 
\appendix
\section{Petrov classification of homogeneous spaces}
In this appendix, we provide in Tables 1-6, the Petrov classification of $S_{\alpha\,\beta}\,,\,C_{\alpha\,\beta}\,,\,\widehat K_{\alpha\,\beta}$ tensors,
for homogeneous spaces of types {\bf A} and {\bf B}, corresponding to unimodular and non-unimodular Lie algebras.

\begin{sidewaystable}
\begin{center}
\begin{tabular}{|c|c|c|c|c|}
\hhline {|=|=|=|=|=|}
 Bianchi  &Discriminant of the traceless Ricci tensor& Special values  & Petrov & Remarks\\
  type {\bf A}& &    &   type &  \\
\hhline {|=|=|=|=|=|}
\multirow{6}{*}{\large$I$}& & & $I_{\mathbb R}$& Generic \\ \cline{3-5}
	&\multirow{4}{*}{ {$\Delta_{\rm S}=-\frac1{27}\,(a+2\,b)^2\,c^2\,\left((a+b)^2-c^2\right)^2\,(a^2-4\,c^2)^2\leqslant0$} }
		&$c=0$,\ $a\neq -b$		&$D_t$	& \\ \cline{3-5}
		 & &$c=0$,\ $a =-b$		&$O$	& $R^\alpha_\beta=-a^2/2\,\delta^\alpha_\beta$,\ $AdS_3$ or flat spaces\\ \cline{3-5}
		& &$ c=\pm (a+b) $		&$D_s$	&  \\   \cline{3-5}
	&
		 &$a=-2\,b$,\ $b\neq c$	&$D_t$	&   If $b=c$, flat space\\ \cline{3-5}
		& &$a=\pm 2\,c$,\ $b\neq \mp c\neq 0$&$D_s$& It $b=\mp c$ or $c=0$, flat space  \\  
		\hhline {|=|=|=|=|=|} 
				\multirow{2}{*}{\large$II$}&\multirow{2}{*}{$\Delta_{\rm S}=0$} & & $II$& Generic \\ \cline{3-5}
					& &$b(a+b)=0$,\ $a \text{ or } b\neq 0 $		&$N$	&If $a=b=0$, flat space \\   
		\hhline {|=|=|=|=|=|} 
				\multirow{2}{*}{\large$III$}&\multirow{2}{*} {$\Delta_{\rm S}=0$} & & $III$& Generic \\ \cline{3-5}	 
		&&$a=0$&$N$ 		&  \\   
		\hhline {|=|=|=|=|=|}
		\multirow{2}{*}{\large$IV$}&\multirow{2}{*} {$\Delta_{\rm S}=\frac{1}{108}(a-b)^2(4\nu^2-a^2)(-ab+b^2+\nu^2)^2(-ab+b^2+4\nu^2)^2\geqslant0,$} & & $I_\mathbb{C}$& Generic \\ \cline{3-5}
		&&$a=b$ 		&$D_s$&  \\    
		\hhline {|=|=|=|=|=|}
		\end{tabular}
\end{center}
\label{SpPnIAS}
\caption{Generic Petrov type of the traceless Ricci tensor on homogeneous space of class $A$ and particular values of the structure tensor density,  
leading to special Petrov types.}
\end{sidewaystable}
\begin{sidewaystable}
\begin{center}
\begin{tabular}{|c|c|c|c|c|}
\hhline {|=|=|=|=|=|}
 Bianchi  &Discriminant of the traceless Ricci tensor & Special values  & Petrov   & Remarks\\
  type {\bf B}& &    &   type &  \\
\hhline {|=|=|=|=|=|}
\multirow{3}{*}{\large$T$}&\multirow{3}{*}{$\Delta_{\rm S}=-\frac{64}{27}\,b^6 \,\left(a^2+k^2\right)\,(a^2+k^2-b^2)^2\leqslant0$} & & $I_{\mathbb R}$& Generic \\ \cline{3-5} 
		   &  &$k^2=b^2-a^2$		&$D_s$	&  \\  \cline{3-5}
		 &&$ b=0 $		&$O$	& $R^\alpha_\beta=2\,k^2\delta^\alpha_\beta$ (locally) $dS_3$   \\   
		 		\hhline {|=|=|=|=|=|} 
\multirow{4}{*}{\large$SI$}&\multirow{4}{*} {$\Delta_{\rm S}=\frac1{108}(a+b)^6\left(4\,k^2-(a-b)^2\right)\,(k^2+a\,b)^2$} & & $I$& $I_{\mathbb C}$ if $4\,k^2>(a-b)^2$, $I_{\mathbb R}$ if $4\,k^2<(a-b)^2$ \\ \cline{3-5}
	& & $k=\pm {(a-b)}/2$& 	$II$	&    $a\neq-b $\\  \cline{3-5}
 		 &&  $a=-b$& $O$& $R^\alpha_\beta=-2\,k^2\delta^\alpha_\beta$ (locally) $AdS_3$   \\  \cline{3-5}
		 	&		
		 &\multirow{2}{*} {$k^2=-a\,b$} &\multirow{2}{*} {$D$}	&$a b<0$; if $|b|>|a|$, type $D_s$,  \\		    
 
		&& & &if $\vert a\vert >\vert b\vert$, type $D_t$\\       
	\hhline {|=|=|=|=|=|} 
				\multirow{2}{*}{\large$SII$}&\multirow{2}{*} {$\Delta_{\rm S}=0$} & & $N$& Generic \\ \cline{3-5}
	 
		&&$k=-a$ &$O$		&$R^\alpha_\beta=-2\,k^2\delta^\alpha_\beta$ (locally) $AdS_3$  \\
				\hhline {|=|=|=|=|=|}
		\multirow{3}{*}{\large$SIII$}&\multirow{3}{*} {$\Delta_{\rm S}=-\frac1{108}\,(4\,k^2-a^2)\,(4\,\nu^2-\,a^2)^3\,(k^2-\,\nu^2)^2,$} & & $I$& $I_{\mathbb R}$ if $k^2>a^2/4$, $k^2\neq \nu^2$, $I_{\mathbb C}$ if $k^2<a^2/4$ \\ \cline{3-5}		 
		 		  &&$k^2=\nu^2$ &$D$ &$D_s$ if $k=\nu$, $D_t$ if $k=-\nu$ \\	 
		 \cline{3-5} 
	  &&$k=\pm a/2$ 		&$II$ &$D_s$ if $a=0$ \\    
		\hhline {|=|=|=|=|=|}
		\multirow{2}{*}{\large$L$}&\multirow{2}{*} {$\Delta_{\rm S}=0$} & & $II$& Generic\\ 
		\cline{3-5}& & $b^2=1+a\,c$ &$D_s$&\\
		\cline{3-5}
	  
		   & & $c=0$		&$N$  & If $b=0$ or $b=-1$, flat space.\\

		 \hhline {|=|=|=|=|=|}
		\end{tabular}
\end{center}
\label{SpPnIBS}
\caption{Generic Petrov type of the traceless Ricci tensor on homogeneous space of class $B$ and particular values of the structure tensor density,  leading to special Petrov types.}
\end{sidewaystable}

\begin{sidewaystable}
\begin{center}

{\small

\begin{tabular}{|c|c|c|c|c|}
\hhline {|=|=|=|=|=|}
 Bianchi  &Discriminant of the Cotton--York tensor  & Special values  & Petrov   & Remarks\\
   type {\bf A}& &   &   type &  \\

 \hhline {|=|=|=|=|=|}
\multirow{6}{*}{\large$I$}&\multirow{6}{*}{$\Delta_{\rm C}=-\frac{1}{1728}c^2((a+b)^2-c^2)^2((3a^2+4c^2)^2-(4ac-8bc)^2)^2\times$} & & $I_{\mathbb R}$& Generic \\ \cline{3-5}
	&\multirow{6}{*}{$\times(a^2-4ab-4(2b^2+c^2))^2\leqslant0$}
		&$c=0$,\ $a\neq -b$		&$D_t$	& \\ \cline{3-5}
		 &&$c=0$,\ $a =-b$		&$O$	& $AdS_3$ or flat spaces\\ \cline{3-5}
		& &$ c=\pm (a+b) $		&$D_s$	&  \\   \cline{3-5}
	&
		 &$c=\pm \frac12\sqrt{(a-2\,b)^2-12\,b^2}$ &$D_t$	&$a\not \in ]-2(\sqrt{3}-1) b,\, 2(\sqrt{3}+1) b[$ 	    \\ \cline{3-5}
		& &$c=\pm\left(\frac a2-b\pm\sqrt{\frac{1}{2}(3\,b^2-(a+b)^2)}~\right)$ &$D_s$&$a  \in ] (\sqrt{3}-1) b,\, -(\sqrt{3}+1) b[$   \\  
		\hhline {|=|=|=|=|=|} 
\multirow{2}{*}{\large$II$}&\multirow{2}{*} {$\Delta_{\rm C}=0$} & & $II$& Generic \\ \cline{3-5}
	 
		&&$b(a+b)=0$,\ $a \text{ or } b\neq 0 $		&$N$	&If $a=b=0$, flat space \\   
		    
	\hhline {|=|=|=|=|=|} 
\multirow{3}{*}{\large$III$}&\multirow{3}{*} {$\Delta_{\rm C}=0  $} & & $III$& Generic \\ \cline{3-5}
	 
		&&$a=0$ 		&$O$&conformally flat,    \\   
	& 	 
		 & & &  Ricci of type $N$ \\ 		\hhline {|=|=|=|=|=|}
\multirow{4}{*}{\large$IV$}&\multirow{4}{*} {$\Delta_{\rm C}=\frac 1{6912}\,\left(4\nu^2-a^2\right)\left((a-b)(3a+b)-4\nu^2\right)^2(-ab+b^2+\nu^2)^2\times $} & & $I_\mathbb{C}$& Generic \\ \cline{3-5}
	&\multirow{4}{*}{$\times \left( -3a^4-8a^3b+2a^2b^2+9b^4+8(a^2+4ab-b^2)\nu^2+16\nu^4\right)^2\geqslant0 $}    &$4\nu^2=(a-b)(3a+b)$ 		&$D_s$ & \\  \cline{3-5}   &&\multirow{2}{*}{$a=-b=\nu$ }& 	$O$	&  conformally flat,    \\ 
	& 	 
		 & & &Ricci of type $I_{\mathbb C}$   \\  
		\hhline {|=|=|=|=|=|}
		\end{tabular}

}		
		
\end{center}

\label{SpPnIAC}
\caption{Generic Petrov type of the Cotton--York tensor on homogeneous space of class $A$ and particular values of the structure tensor density,  leading to special Petrov types.}
\end{sidewaystable}

\begin{sidewaystable}
\begin{center}

{\small

\begin{tabular}{|c|c|c|c|c|}
\hhline {|=|=|=|=|=|}
 Bianchi  &Discriminant of the Cotton--York tensor& Special values  & Petrov   & Remarks\\
   type {\bf B}& &   &   type & \\

 \hhline {|=|=|=|=|=|}
\multirow{6}{*}{\large$T$}&\multirow{6}{*} {$\Delta_{\rm C}= -\frac{64}{27} b^6 \left(a^2-b^2+k^2\right)^2 \left(4 a^2-b^2+k^2\right)^2 \times $} & &$I_{\mathbb R}$& 
Generic \\ \cline{3-5}
	& \multirow{6}{*}{$\times \left(\left(2 a^2+b^2\right)^2+k^2 \left(5 a^2-2
   b^2\right)+k^4\right)\leqslant0$}
		&$b=0$&$O$&$dS_3$ \\ \cline{3-5}
		&&$k^2=b^2-a^2$&$D_s$& \\ \cline{3-5}
		&&$k^2=b^2-4\,a^2$&$D_s$&  \\   \cline{3-5}
	&
		 &$k^2=b^2$, $a=0$&$O$& Conformally flat, Ricci type $D_s$	    \\   
		\hhline {|=|=|=|=|=|} 
\multirow{6}{*}{\large$SI$}&\multirow{6}{*} {$\Delta_{\rm C}=  -\frac 1 {6912}{(a+b)^6\left(a b+k^2\right)^2 \left((a-3 b) (3 a-b)-4 k^2\right)^2  } \times $} & & $I$&If $(...)>0$ $I_{\mathbb R}$ and  if $(...)<0$ $I_{\mathbb C}$ \\ \cline{3-5}
	& \multirow{6}{*}{$\left(\left(3(a-b)^2+4ab\right)^2-4 k^2 \left(3(a-b)^2-8ab\right)+16 k^4\right)$}  		&$a=-b$&$O$&$AdS_3$ \\ \cline{3-5}
		&&$k^2=-a\,b$&$D$ &$a\,b<0$, $D_s$ if $\vert b\vert>\vert a \vert$, $D_t$ if $\vert a\vert>\vert  b\vert$ \\ \cline{3-5}
		&&$4k^2=(a-3\,b)(3\a-b)$&$D$ &$a\not \in [b/3,3\,b]$, $D_s$ if $a\in [-b,3\,b]$, $D_t$ if $a\not\in [-b,\,3\,b]$
		 		 \\   \cline{3-5}
	&&$16k^2=3(a-b)\pm\sqrt{-(3a+b)(a+3b)}$&\multirow{1}{*}{ $II$}&\multirow{1}{*}{ $O$ if $b=-a$ and $k=a^2$ or $4a^2$}	  \\ 
		\hhline {|=|=|=|=|=|}
\multirow{3}{*} {\large$SII$}&\multirow{3}{*}{$\Delta_{\rm C}=0$} & & { $N$}& { Generic} \\  \cline{3-5}
&&$a=-k/2$&$O$&Ricci type $N$\\ \cline{3-5}
&&$a=-k$&$O$& $AdS_3$\\
 		\hhline {|=|=|=|=|=|}
 \multirow{4}{*}{\large$SIII$}& \multirow{4}{*}{$\Delta_{\rm C}= \frac 1{6912} \left(4 \nu ^2-a^2\right)^3 \left(k^2-\nu ^2\right)^2 \left(4\nu^2-4k^2+3\,a^2\right)^2\times $} & &$I$&If $(...)<0$ $I_{\mathbb R}$ and  if $(...)>0$ $I_{\mathbb C}$ \\ \cline{3-5}
	&  \multirow{4}{*}{$\times \left(9a^4+16(k^2-\nu^2)^2-12a^2(k^2+2\nu^2)\right)$}		&$k^2=\nu^2$&$D $&$D_s$ if $k=\nu$, $D_t$ if $k=-\nu$ \\ \cline{3-5}
				&&$4k^2=4\,\nu^2+3\,a^2$& $D$&$D_s$ if $\nu\,k>0$, $D_t$ if $\nu\,k<0$ \\   \cline{3-5}
				&& &&\\ 
	& 
		 &$16k^2=3a\pm\sqrt{16\nu^2-3a^2} $&$II$& $O$ if $a=0$ and $k^2=\nu^2$   \\  
		\hhline {|=|=|=|=|=|} 
\multirow{3}{*}{\large$L$}&\multirow{3}{*}{  $\Delta_{\rm C}= 
0$} & & $II$& Generic \\ 
\cline{3-5}& & $b^2=1+a\,c$ &$D_s$&\\
\cline{3-5}		&&$c=0$&$O$&Conformally flat, \\ 
		  	& 		  &&&Ricci type $N$	    \\ 	 
		\hhline {|=|=|=|=|=|} 
		\end{tabular}

}		
		
\end{center}

\label{SpPnIBC}
\caption{Generic Petrov type of the Cotton--York tensor on homogeneous space of class $B$ and particular values of the structure tensor density,  leading to special Petrov types.}
\end{sidewaystable}

\begin{sidewaystable}
\begin{center}

{\small

\begin{tabular}{|c|c|c|c|c|}
\hhline {|=|=|=|=|=|}
 Bianchi  &Discriminant of the $\widehat K_{\alpha\,\beta}$ tensor  & Special values  & Petrov   & Remarks\\
   type {\bf A}& &   &   type &  \\

 \hhline {|=|=|=|=|=|}
\multirow{6}{*}{\large$I$}&\multirow{6}{*}{$\Delta_{\rm \widehat K}=\cdots\leqslant0$, in  \eqn{DKI}} & & $I_{\mathbb R}$& Generic \\ \cline{3-5}
	&\multirow{6}{*}{}
		&$c=0$,\ $(4b+21a)(b+a)\neq 0$		&$D_t$	& \\ \cline{3-5}
		 &&$c=0$,\ $(4b+21a)(b+a)=0$		&$O$	& for $b=-a$ $AdS_3$ or flat spaces\\ \cline{3-5}
		& &$ c=\pm (a+b) $		&$D_s$	&  \\   \cline{3-5}
	&
		 &$c=\pm\frac{\sqrt{-5a^2+2a^2b-40ab^2-64b^3}}{2\sqrt{5a+26b}} $ &$D_t$	& $O:~(a+2b)(25a+42b)(5a^3+36a^2b-44ab^2+64b^3)=0$	    \\ \cline{3-5}
		& &$21a^3+4ac(\pm6b+5c)+a^2(4b\pm26c)\pm8c(8b^2\pm8bc+5c^2)=0$ &$D_s$&   \\  
		\hhline {|=|=|=|=|=|} 
\multirow{2}{*}{\large$II$}&\multirow{2}{*} {$\Delta_{\rm\widehat K}=0$} & & $II$& Generic \\ \cline{3-5}
	 
		&&$b(4b+21a)(b+a)=0$,\ $a \text{ or } b\neq 0 $		&$N$	&If $a=b=0$, flat space \\   
		    
	\hhline {|=|=|=|=|=|} 
\multirow{4}{*}{\large$III$}&\multirow{3}{*} {$\Delta_{\rm\widehat K}=0  $} & & $III$& Generic \\ \cline{3-5}
	 
		&&$a=0$ 		&$O$&conformally flat,    \\   
	& 	 
		 & & &  Ricci of type $N$  \\ 		\hhline {|=|=|=|=|=|}
\multirow{4}{*}{\large$IV$}&\multirow{4}{*} {$\Delta_{\rm\widehat K}=\cdots \geqslant0$, in  \eqn{DKIV}} & & $I_\mathbb{C}$& Generic \\ \cline{3-5}
	&   &$-21a^3+15a^2b+ab^2+5b^3+4(13a-5b)\nu^2=0$ 		&$D_s$ & \\   &&\multirow{2}{*}{}&&    \\ 
		\hhline {|=|=|=|=|=|}
		\end{tabular}

}		
		
\end{center}

\label{SpPnIAK}
\caption{Generic Petrov type of the $\widehat K_{\alpha\,\beta}$ tensor on homogeneous space of class $A$ and particular values of the structure tensor density,  leading to special Petrov types.}
\end{sidewaystable}

\begin{sidewaystable}
\begin{center}

{\small

\begin{tabular}{|c|c|c|c|c|}
\hhline {|=|=|=|=|=|}
 Bianchi  &Discriminant of the $\widehat K_{\alpha\,\beta}$ tensor& Special values  & Petrov   & Remarks\\
   type {\bf B}& &   &   type & \\

 \hhline {|=|=|=|=|=|}
\multirow{6}{*}{\large$T$}&\multirow{6}{*} {$\Delta_{\rm\widehat K}=\cdots\leqslant0$,  in  \eqn{DKT}} & &$I_{\mathbb R}$& Generic \\ \cline{3-5}
	& \multirow{6}{*}{}
		&$b=0$&$O$&$dS_3$ \\ \cline{3-5}
		&&$k^2=b^2-a^2$&$D_s$& \\ \cline{3-5}
	&
		 &$k^2=-24a^2+5b^2\pm16\sqrt{2}\vert a\vert\sqrt{a^2-b^2}$&$D_s$& 	     \\
		\hhline {|=|=|=|=|=|} 
\multirow{4}{*}{\large$SI$}&\multirow{4}{*} {$\Delta_{\rm\widehat K}= -\#\left((a-b)^2(21a^2+10ab+21b^2)^2-\right.$} & & $I$&If $(...)>0$ $I_{\mathbb R}$ and  if $(...)<0$ $I_{\mathbb C}$ \\ \cline{3-5}
	& \multirow{4}{*}{$-4(63a^4-348a^3b+970a^2b^2-348ab^3+63b^4)k^2$}  		&$a=-b$&$O$&$AdS_3$ \\ \cline{3-5}
		& \multirow{4}{*}{$\left.+16(39a^2-118ab+38b^2)k^4-64k^6\right)~,\#\geqslant0$, in  \eqn{DKSI}}&$k^2=-a\,b$&$D$ &$a\,b<0$, $D_s$ if $\vert b\vert>\vert a \vert$ and $D_t$ if $\vert a\vert>\vert  b\vert$ \\ \cline{3-5}
		&&$4k^2=19(a-b)^2-20ab\pm16\vert a-b\vert\sqrt{a^2-6ab+b^2}$&$D_{t,s}$ & $(a+b)\left(5a-3b\pm{\rm s}[a-b]\sqrt{a^2+b^2-6ab}~\right)\gtrless 0$
		 		 \\   \cline{3-5}
	&&$(\cdots)=0$&$D_t$&	  \\ 
		\hhline {|=|=|=|=|=|}
\multirow{3}{*} {\large$SII$}&\multirow{3}{*}{$\Delta_{\rm\widehat K}=0$} & & { $N$}& { Generic} \\  \cline{3-5}
&&$4a=k(-2\pm\sqrt{2}~)$&$O$&Ricci type $N$\\ \cline{3-5}
&&$a=-k$&$O$& $AdS_3$\\
 		\hhline {|=|=|=|=|=|}
 \multirow{4}{*}{\large$SIII$}& \multirow{4}{*}{$\Delta_{\rm\widehat K}=\#\left(441a^6-84a^4(3k^2+26\nu^2)-64(k^3-5k\nu^2)^2+\right.$} & &$I$&If $(...)<0$ $I_{\mathbb R}$ and  if $(...)>0$ $I_{\mathbb C}$ \\ \cline{3-5}
	&  \multirow{4}{*}{$\left.+16a^2(39k^4-24k^2\nu^2+169\nu^4)\right)~,\#\geqslant0$,  in  \eqn{DKSIII}}		&$k^2=\nu^2$&$D $&$D_s$ if $k=\nu$, $D_t$ if $k=-\nu$ \\ \cline{3-5}
				&&$4k^2=19a^2+20\nu^2\pm16\vert a\vert\sqrt{a^2+4\nu^2}$& $D$&$D_s$ if $\nu\,k>0$, $D_t$ if $\nu\,k<0$ \\   \cline{3-5}
	& 
		 &$(\cdots)=0$&$II$& $O$ if $a=0$ and $k^2=5\nu^2$   \\  
		\hhline {|=|=|=|=|=|} 
\multirow{3}{*}{\large$L$}&\multirow{3}{*}{  $\Delta_{\rm\widehat K}= 
0$} & & $II$& Generic \\
\cline{3-5}& & $b^2=1+a\,c$ &$D_s$&\\
\cline{3-5}		&&$c=0$&$O$&Conformally flat, Ricci type $N$\\ 	 
		\hhline {|=|=|=|=|=|} 
		\end{tabular}

}		
		
\end{center}

\label{SpPnIBK}
\caption{Generic Petrov type of the $\widehat K_{\alpha\,\beta}$ tensor on homogeneous space of class $B$ and particular values of the structure tensor density,  
leading to special Petrov types.}
\end{sidewaystable}

\clearpage

Finally, we give the expressions of the discriminants of $\widehat K_{\alpha\,\beta}$ tensor for all the cases
\begin{eqnarray}
\label{DKI}
&&\Delta^I_{\widehat K}=-\frac{1}{110592}((a +b)^2 -c^2 )^2 c ^2 \left(5 a ^3-2 a ^2 b +40 a  b ^2+64 b ^3+4 (5 a +26 b ) c ^2\right)^2\times\nonumber \\
&&\times\left(21 a ^3+4 a  c  (-6 b +5 c )+a ^2 (4 b +26 c )+8 c  \left(8 b ^2-8 b  c 
+5 c ^2\right)\right)^2\times\nonumber \\
 &&\times\left(21 a ^3+a ^2 (4 b -26 c )+4 a  c  (6 b +5 c )
 -8 c  \left(8 b ^2+8 b  c +5 c ^2\right)\right)^2\leqslant0\qquad,\\
 &&\Delta^{II}_{\widehat K}=0\qquad,\\
&& \Delta^{III}_{\widehat K}=0\qquad,\\
 \label{DKIV}
 &&\Delta^{IV}_{\widehat K}=\frac{1}{442368}\left(4 \nu ^2-a ^2\right) \left(-a  b +b ^2+\nu ^2\right)^2 \left(-21 a ^3+15 a ^2 b +a  b ^2+5 b ^3+4 (13 a -5 b ) \nu ^2\right)^2\times\nonumber\\ 
&& \left(105 a ^6+76 a ^5 b +381 a ^4 b ^2-160 a ^3 b ^3-45 a ^2 b ^4+84 a  b ^5-441 b ^6-
4 \left(171 a ^4+56 a ^3 b +486 a ^2 b ^2-160 a  b ^3-\right.\right.\nonumber\\
&&\left.\left.-41 b ^4\right) \nu ^2+16 \left(91 a ^2-20 a  b +105 b ^2\right) \nu ^4-1600 \nu ^6\right)^2\geqslant0\qquad,\\
\label{DKT}
&&\Delta^T_{\widehat K}=-\frac{64}{27} b ^6 \left(a ^2-b ^2+k ^2\right)^2 \left(\left(8 a ^3+13 a  b ^2\right)^2
+\left(112 a ^4-74 a ^2 b ^2+25 b ^4\right) k ^2+\left(49 a ^2-10 b ^2\right) k ^4+k ^6\right)\times\nonumber\\ 
&&\times\left(64 a ^4+\left(-5 b ^2+k ^2\right)^2+16 a ^2 \left(b ^2+3 k ^2\right)\right)^2\leqslant0\qquad,\\
\label{DKSI}
&&\Delta_{\rm\widehat K}^{SI}=-\frac{1}{442368}(a +b )^6 \left(a  b +k ^2\right)^2 \left(105 a ^4-156 a ^3 b +502 a ^2 b ^2-156 a  b ^3+105 b ^4-8 \left(19 a ^2-58 a  b +19 b ^2\right) k ^2+\right.\nonumber\\
&&\left.+16 k ^4\right)^2 \left((a -b )^2 \left(21 a ^2+10 a  b +21 b ^2\right)^2-4 \left(63 a ^4-348 a ^3 b +970 a ^2 b ^2-348 a  b ^3+63 b ^4\right) k ^2+\right.\nonumber\\
&&\left.16 \left(39 a ^2-118 a  b +39 b ^2\right) k ^4-64 k ^6\right)\qquad,\\
&&\Delta_{\rm\widehat K}^{SII}=0\qquad,\\
\label{DKSIII}
&&\Delta_{\rm\widehat K}^{SIII}=\frac{1}{442368}\left(4 \nu ^2-a^2\right)^3 \left(k ^2-\nu ^2\right)^2 \left(105 a ^4+16 \left(k ^2-5 \nu ^2\right)^2-8 a ^2 \left(19 k ^2+33 \nu ^2\right)\right)^2\times\nonumber\\
&&\times\left(441 a ^6-84 a ^4 \left(3 k ^2+26 \nu ^2\right)-64 \left(k ^3-5 k  \nu ^2\right)^2+16 a ^2 \left(39 k ^4-24 k ^2 \nu ^2+169 \nu ^4\right)\right)\qquad,\\
&&\Delta^L_{\widehat K}=0\qquad.
\end{eqnarray}

\end{document}